\documentclass{article}

\usepackage[english]{babel}
\usepackage[letterpaper,top=2cm,bottom=2cm,left=3cm,right=3cm,marginparwidth=1.75cm]{geometry}
\usepackage{mathptmx}
\usepackage{sectsty}
\usepackage{amsmath}
\usepackage{bm}
\usepackage{graphicx} 
\usepackage{hyperref}
\usepackage[format=plain,justification=justified,singlelinecheck=false,font={stretch=1.125,small,sf},labelfont=bf,labelsep=space]{caption}

\title{Dynamics of Ball-Chains and Very Elastic Fibres Settling under Gravity in a Viscous Fluid$^\dag$}

\author{H. J. Shashank,\textit{$^{a}$} Yevgen Melikhov,\textit{$^{a}$} and Maria L. Ekiel-Je\.zewska$^{\ast}$\textit{$^{a}$} }

\begin{document}

\renewcommand{\thefootnote}{\fnsymbol{footnote}}
\renewcommand\footnoterule{\vspace*{1pt}
 \hrule width 3.5in height 0.4pt \vspace*{5pt}} 

\maketitle

\begin{abstract}
    We study experimentally the dynamics of one and two ball-chains settling under gravity in a very viscous fluid at a Reynolds number much smaller than unity. We demonstrate that single ball-chains in most cases do not tend to be planar and often rotate, not keeping the ends at the same horizontal level. Shorter ball-chains usually form shapes resembling distorted U, and longer ones in the early stage of the evolution form a shape resembling distorted W, and later deform non-symmetrically and significantly out of plane. This behaviour is reproduced in our numerical simulations for a single very elastic filament, with the use of the bead model and multipole expansion of the Stokes equations, corrected for lubrication and implemented in the precise Hydromultipole numerical codes. In our experiments, two ball-chains, initially one above the other, later move away or approach each other, for a larger or smaller initial distance, respectively.
\end{abstract}

\footnotetext{\textit{$^{a}$}~Institute of Fundamental Technological Research, Polish Academy of Sciences, ul. Pawi\'nskiego 5B, 02-106 Warsaw, Poland}

\footnotetext{$^{\ast}$~corresponding author: Maria L. Ekiel-Je\.zewska E-mail: mekiel@ippt.pan.pl}

\footnotetext{\dag~Electronic Supplementary Information (ESI) is available \label{reftoESI} }

\section{Introduction} \label{Intro}
Recently, there has been a lot of interest in studying the motion of flexible and rigid microfibres under external forces \cite{witten2020review} or ambient flows \cite{du2019dynamics,Linder2015}. This research is guided by many potential applications %
to biological systems \cite{nguyen2014hydrodynamics,kantsler2012}, to medical diagnostic techniques \cite{zhang2020flow} or %
in to design of new materials \cite{nunes2013,perazzo2017}. 
In particular, it is important to study the effect of gravity on swimming or just settling deformable microorganisms \cite{slomka2020encounter,rizvi2020deformable}, and on flexible microobjects produced by innovative modern technologies \cite{xue2022shear}.
Therefore, sedimentation of single or multiple deformable objects of different types has been investigated for a wide range of bending stiffness, and for different shapes \cite{saggiorato2015conformations,bukowicki2018different,yu2021coil,silmore2021buckling}, also for non-negligible inertia effects \cite{ravichandran2022orientation}.

Dynamics of one, two, or three elastic fibres settling under gravity in a viscous fluid at a Reynolds number much smaller than unity has been extensively investigated theoretically and numerically \cite{xu1994deformation,lagomarsino2005hydrodynamic,schlagberger2005orientation,manghi2006,li2013sedimentation,saggiorato2015conformations,shojaei2015sedimentation,marchetti2018deformation, bukowicki2018different,bukowicki2019sedimenting,koshakji2023robust}. %
Different types of motion, hydrodynamic repulsion or attraction, and shape deformations have been found, depending on values of the so-called elasto-gravitational number, equal to the ratio of gravitational to elastic forces, and also depending on initial relative orientations and relative positions of the fibres. %

However, the number of experimental studies of sedimenting flexible fibres, as far as we know,  %
has been much smaller, and limited to rather short and moderately elastic fibres that tend to reach a stable, stationary, \textquotesingle V\textquotesingle{} or \textquotesingle U\textquotesingle{} shaped configuration \cite{marchetti2018deformation}.

Therefore, the main goal of this paper is to explore experimentally the dynamics and shape evolution of very flexible elongated objects. After many trials, we decided to focus on investigating a single ball-chain or two ball-chains close to each other that settle %
under gravity in a viscous fluid at a low Reynolds number. %
It is expected that the ball-chain dynamics is similar to those of %
very elastic filaments, with the number of beads on the 
object determining its ability to bend, as 
the bending of each triplet of the consecutive beads is limited by geometry. 
Using ball-chains with a moderate or relatively large number of beads, we can compare their observed behaviour with the dynamics predicted numerically for very flexible elastic filaments \cite{saggiorato2015conformations}, which has not been done experimentally up until now. 

In general, we aim towards extracting experimentally basic features of the dynamics of very flexible elongated objects. Next, we will perform numerical simulations to determine these features using a theoretical description. One of the interesting goals is to investigate how a planar vertical symmetric U-shape, inherent for moderately flexible filaments, becomes unstable when their flexibility is increased. 

The paper is structured as follows. In Section 2, we present the experimental setup, materials and the methods used, including the image processing techniques. Section 3 contains the experimental results - evolution of shapes and positions of a single ball-chain and of a pair of ball-chains, illustrated also in the supplementary videos. Then, in Section 4, we present numerical simulations for a single elastic filament - first, the theoretical description and its numerical implementation {\sc Hydromultipole}, and next, the numerical results that agree with the experiments remarkably well. Finally, we conclude in Section 5.

\section{Experimental Techniques, Materials \& Methods} \label{Exp_techniques_materials_methods}

\subsection{Experimental Arrangement} \label{Exp_setup}

We conduct experiments within a glass tank of inner dimensions $200 mm$ width, %
$200 mm$ depth and %
$500 mm$ height, that is filled with highly viscous silicon oil (manufactured by \textit{Silikony Polskie}) with kinematic viscosity $\nu = 5 \times 10^{-3} m^2/s$ and density $\rho$ = 970 $kg/m^3$ at 25$^{\circ}C$. Into this tank, we drop flexible ball-chains.

\begin{figure}%
 \centering
 \includegraphics[width=0.4\textwidth]{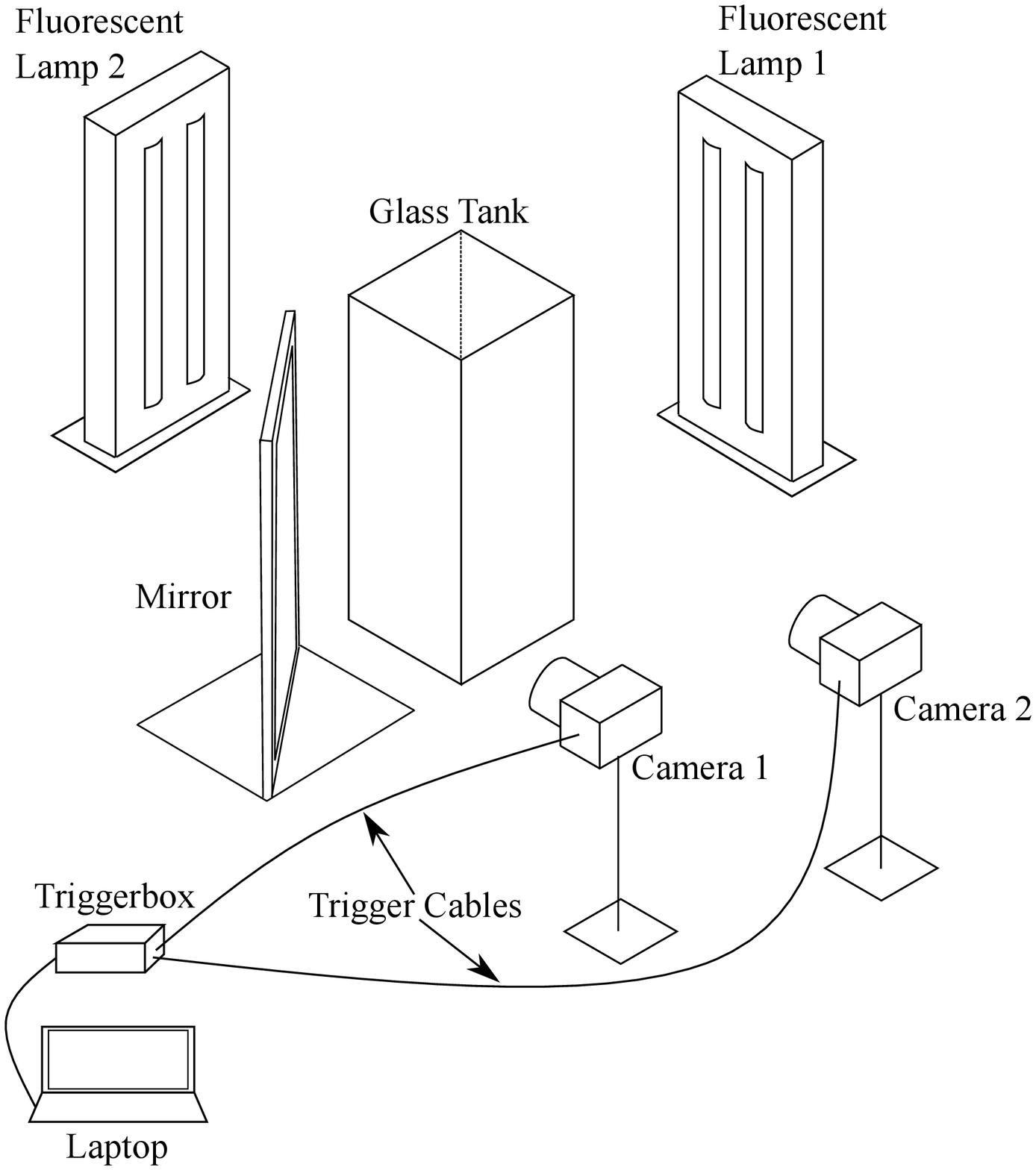}
 \caption{Schematic of %
 the experimental arrangement.}
 \label{fig:figure1}
\end{figure}

The motion of the ball-chains is viewed using two cameras placed in perpendicular orientations, corresponding to the front and side views, as shown in the schematic of the experimental setup in Fig.~\ref{fig:figure1}. Camera 2 views the glass tank directly, while Camera 1 views the glass tank via a first surface mirror (manufactured by \textit{First Surface Mirror LLC}, USA) placed at an angle of 45$^{\circ}$. These two views provide a greater insight into the lateral movements of the ball-chains. A fluorescent lamp is placed behind each of the vertical faces of the tank opposite to where the cameras are located. Camera 1 is illuminated by Fluorescent lamp 1, through the mirror, and Camera 2 is illuminated by Fluorescent lamp 2, as seen in the schematic shown in Fig.~\ref{fig:figure1}. This arrangement of the lamps ensures the best illumination of the tank, with the opaque sedimenting ball-chains being clearly contrasted from the bright background.

The cameras used are two identical full-frame DSLRs (Canon 5D Mark iv) with a resolution of 30 megapixels and equipped with a $100 mm$ prime lens. Both cameras are triggered externally so that photographs can be captured at the same time. We use an \textit{Esper Triggerbox} to trigger both cameras simultaneously. The triggerbox maintains a trigger delay of less than $20 ms$ between the two cameras. The triggerbox itself is controlled by a laptop, and the photographs are captured at the maximum permissible rate allowed by the DSLR cameras (1 photo per second). The exposure time of the photographs from both cameras is set to $1/125 s$ to ensure that the motion of the ball-chains remains frozen for the entire duration of the exposure time. The distance travelled by the ball-chains in this time is smaller than a pixel. The f-number is set to its highest available, f/32 (i.e. the smallest aperture), to ensure that the ball-chains remain in focus even as they meander out of the focal plane of the cameras. Finally, the ISO rating is kept at 400 to ensure sufficient image brightness, whilst also ensuring a low to moderate noise levels. 

The cameras are placed in portrait orientation at about $800 mm$ from the front faces of the glass tank, %
according to their respective views. This gives a field of view of about $300 mm$ in the vertical direction and $200 mm$ in the horizontal direction. We centre the cameras at the middle of the glass tank so that approximately $100 mm$ from the free surface and $100 mm$ from the bottom glass face are absent in the photographs, thereby reducing the effect of the free surface and the bottom glass wall on the sedimenting ball-chains.

\subsection{Ball-Chains} \label{Ball_chains}
The ball-chains that are used in this work (manufactured by \textit{Koniarscy S.C.}) %
consist of metallic beads connected to each other through an inextensible string, 
with a possibility of a slight movement of the beads with respect to the string.

The diameter of all metallic beads is the same, $d = 1.5 mm$, and ball-chains of different lengths (i.e., different numbers of beads $N$) have been used, typically with $N$=12 and $N$=20. The filament length $L$ is defined as  
$L = Nd + (N-1)d_b$, where $d_b$ is the length of a string that connects two consecutive beads. In our experiments, 
$d_b \approx 0.3 mm$ 
(on average, since the beads can slightly slide over the string).

\begin{figure}[h!]
 \centering
 \includegraphics[width=0.31\textwidth]{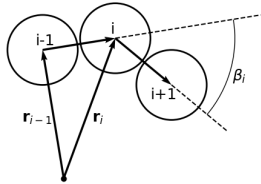}
 \caption{Schematic of a section of a ball-chain. A local bending angle $\beta_i$ is shown.} %
 \label{fig:angle_beta}
\end{figure}

The ball-chains are more dense than silicon oil and settle down owing to gravity. Hydrodynamic interactions between the beads cause the whole ball-chain to bend while moving. The ball-chains do not have an inherent elasticity, and this allows them to bend with zero spending of energy. Bending angle $\beta_i$ of a triplet of consecutive beads, $i-1$, $i$ and $i+1$, is defined by the relation
\begin{equation}
\label{eq:beta}
    \cos \beta_i=\frac{(\bm{r}_i-\bm{r}_{i-1})\cdot (\bm{r}_{i+1}-\bm{r}_i)}{|\bm{r}_i-\bm{r}_{i-1}| |\bm{r}_{i+1}-\bm{r}_i|},
\end{equation}
where $\bm{r}_i=(x_i,y_i,z_i)$ is the position of the centre of the bead $i$ (see Fig.~\ref{fig:angle_beta}).

The change of shape %
of a ball-chain is limited %
by: (a) the maximum bending angle of a bead triplet, %
and (b) the number of beads in the ball-chain.  Naturally, a ball-chain that allows a larger %
bending angle of a bead triplet %
would allow for %
larger bending of the whole %
ball-chain. Furthermore,
for a larger %
number of beads in the ball-chain  %
the shape can be more bent, %
too. %
We have measured the bending angles of bead triplets of the sedimenting ball-chains, as well as when the ball-chain is outside the fluid, free of gravitational effects. In the latter case, we find that the maximum bending angle $\beta_i$ of a bead triplet %
is 55$^{\circ}$ when we force the ball-chain to bend to its maximum possible extent. On the other hand, the maximum bending angle of a bead triplet of a sedimenting ball-chain ranges from 33$^{\circ}$ to 40$^{\circ}$ indicating that the ball-chain does not bend to its full extent as it sediments in our experiments.    

In this paper, we will study %
sedimentation of a single type of a ball-chain, thus ensuring that the ball-chain maximum deformation %
is a function of the number of beads alone. We will first investigate %
time-dependent shape and velocity of a single ball-chain settling under gravity in the silicon oil.  %
We will then present the interaction between two ball-chains that sediment close to each other, but do not touch. %

\subsection{Experimental Methods \& Analysis} \label{Exp_methods}
The experiments are conducted by manually dropping one or two ball-chains approximately at the centre of the tank,  in an approximately horizontal orientation. In the case of single ball-chain runs, a ball-chain is always inserted parallel to the plane of view of camera 2. The experiments with two ball-chains sedimenting close to each other are performed by placing the ball-chains one by one, at a certain time separation, resulting in the position one above the other in a perpendicular orientation. The bottom ball-chain is always placed earlier and parallel to the plane of view of camera 2 (perpendicular to the plane of view of camera 1) and the top ball-chain is always placed later and parallel to the plane of view of camera 1 (perpendicular to the plane of view of camera 2). The cameras are triggered at the moment when both ball-chains are already %
inside the tank, and photographs are acquired until the top ball-chain exits the field of view. We stress that, given the pliable nature of the ball-chains, it is not %
possible to ensure that the ball-chains are always exactly horizontal nor exactly straight once they are inserted. Moreover, the ball-chains %
bend transversely %
just after
entering the fluid, and they are already significantly curved when they enter the camera field of view.

\begin{figure}[h!]
    \centering \vspace{-.4cm}
    \includegraphics[width=0.35\textwidth]{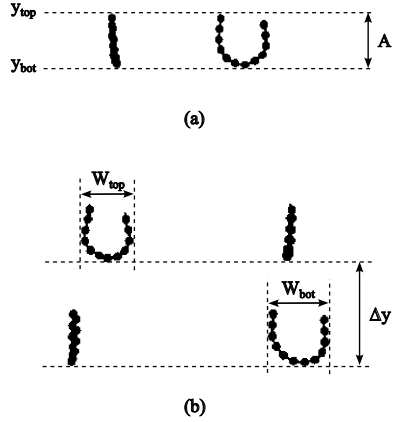}
    \vspace{-.4cm}
    \caption{
    Schematic describing %
    definition of parameters. 
    (a) Bending amplitude $A$, determined in 
    single and multiple ball-chain experiments.
    (b) Vertical distance $\Delta y$ between
    two ball-chains and their widths $W$.
    }
    \label{fig:figure1a}
\end{figure}

We analyse the motion of the ball-chains by extracting relevant data from the photographic image sequence obtained from both cameras. The recorded photographs from each camera are imported into MATLAB, and image processing techniques are applied %
to identify each of the ball-chains. The post-processing steps involve image thresholding, image binarisation and performing morphological operations on the binarised image. Following these steps, each ball-chain is uniquely identified in each frame and in each camera view. Once the ball-chains are uniquely identified, we then calculate the various parameters of the ball-chains to better understand their sedimentation mechanics. Some of the characteristic parameters are illustrated in Fig.~\ref{fig:figure1a}. %

(i) Following Refs. \cite{lagomarsino2005hydrodynamic,bukowicki2018different}, we determine the bending amplitude $A$ as the absolute difference between the %
uppermost location $y_{top}$ and the %
lowermost location $y_{bot}$ of the ball-chain, $A = |y_{top} - y_{bot}|$, and we divide it by the filament length $L$ (defined in the previous section). %

(ii) We also evaluate the vertical component $v_y$ of a ball-chain time-dependent velocity as the ratio of the vertical distance 
between its %
lowermost locations at the times of the two consecutive photographs (approximately equal to the distance travelled by the ball-chain between two consecutive photographs) 
to the time difference between those photographs, $v_y = \frac{y_{bot}^{t + \Delta_t} - y_{bot}^{t}}{\Delta t}$, with $\Delta t = 1 s$. 
These quantities are first calculated separately from each camera view, and are later averaged across the two views.

The fluid and the ball-chains are chosen in such a way that a typical Reynolds number in the experiments is much smaller than unity, with 
$Re=7 \cdot 10^{-4}-10^{-3}$ 
if based on the ball-chain width, and 
$Re=8 \cdot 10^{-3}-2\cdot 10^{-2}$ 
if based on the ball-chain length. 

We will now discuss the uncertainty of the measurements. 
The accuracy in the calculation of $A$ and $v_y$ depends on the accuracy of the identification of $y_{bot}$ and $y_{top}$ during post-processing. The high contrast %
between the back-lit image %
of the ball-chains and the bright background %
ensure that the errors in identifying the edges are minimal. The error of $y_{bot}$ and $y_{top}$ is around $\pm$ 0.1 $mm$ ($\pm$ 2 pixels), which leads to the bending amplitude $A$ having an uncertainty of $\pm$ 0.15 $mm$, and the vertical velocity $v_y$ an uncertainty of $\pm$ 0.19 $mm/s$. This yields a measurement uncertainty of about 10\% for a single camera measurement. Moreover, during the combined measurement with both cameras, the proximity of cameras to the glass tank, coupled with the tendency of the ball-chains to drift off the focal plane of the cameras, %
may cause a systematic %
error due to a parallax. The parallax errors are most noticeable at the extreme ends of the field of view, \textit{i.e.}, at the top and bottom. %
However, the influence of parallax is not significant for the measurement of local quantities, such as the bending amplitude %
$A$ and the local velocity $v_y$. %
Furthermore, the side walls %
also affect the dynamics. Long-range hydrodynamic interactions of the ball-chains with the walls slow down the motion and may influence their shapes, particularly if the ball-chains drift away from the centre of the tank.

\section{Experimental %
Results} \label{Exp_results}
\subsection{Sedimentation of a Single Ball-Chain} \label{single_ballchain}

\begin{figure*}[t!]
    \centering
    \includegraphics[width = 0.95\textwidth]{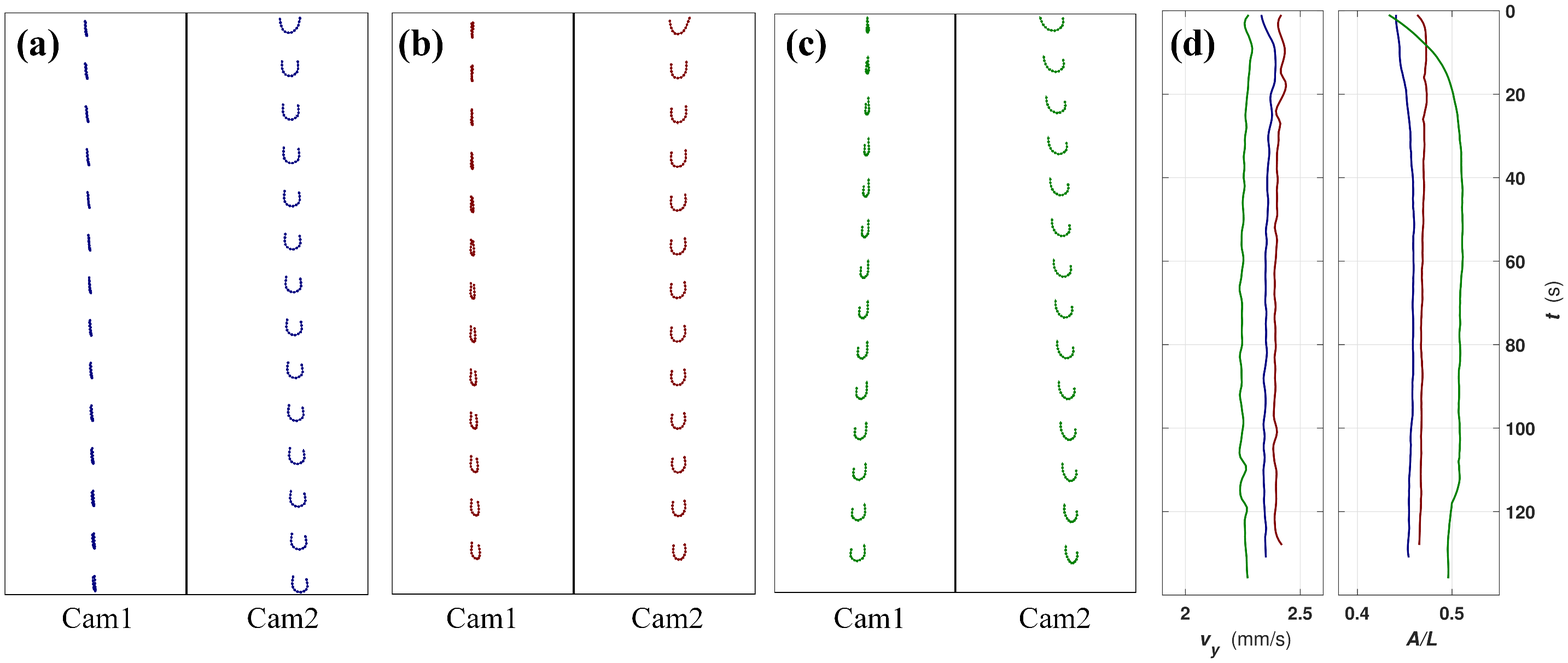}\vspace{-.05cm}
    \caption{Snapshots of a single 12-bead ball-chain settling under gravity in a viscous oil, taken simultaneously by two cameras (Cam1 and Cam2) located at the same level, and with perpendicular lines of sight: %
    (a) a trial without rotation; %
    (b) a trial with rotation; %
    (c) a trial with rotation and non-horizontal orientation of the end-to-end vector; %
    (d) %
    the sedimentation velocity $v_y$ and the bending amplitude $A/L$ vs. time $t$, plotted with blue, red and green lines for the trials shown in (a),  (b) and (c), respectively.
    Experimental movies that correspond to the trials (a), (b) and (c) are shown in the ESI\dag{} in Videos 4a, 4b and 4c, respectively.
    }
    \label{fig:figure2}
\end{figure*}

\begin{figure*}[t!]
    \centering
    \includegraphics[width = 0.95\textwidth]{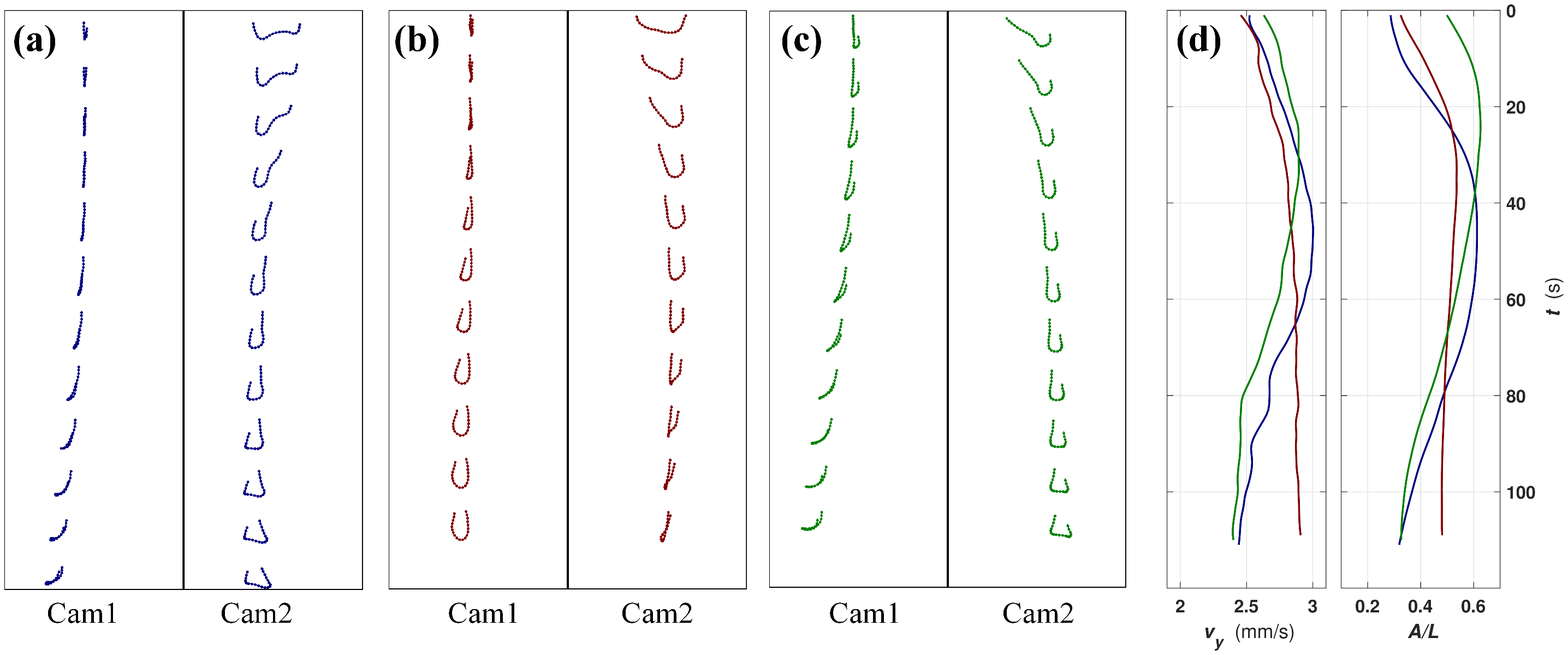}\vspace{-.05cm}
    \caption{Snapshots of a single 20-bead ball-chain settling under gravity in a viscous oil, taken simultaneously by two cameras (Cam1 and Cam2) located at the same level, and with perpendicular lines of sight: %
    (a) a trial that begins with a W-shape; %
    (b) a trial that begins with a U-shape; %
    (c) a trial that begins with an asymmetric hook-shape;
    (d) %
    the sedimentation velocity $v_y$ and the bending amplitude $A/L$ vs. time, plotted with blue, red and green lines for the trials shown in (a),  (b) and (c), respectively.
    Experimental movies that correspond to the trials (a), (b) and (c) are shown in the ESI\dag{} in Videos 5a, 5b and 5c, respectively.
    }
    \label{fig:figure3}
\end{figure*}

We recorded and analysed sedimentation of a single ball-chain in 5 %
experimental trials for 12-bead ball-chains and in 14 %
experimental trials for 20-bead ball-chains. (The shorter ball-chains did not show rich dynamics in contrast to the longer ones, hence the moderate difference in the number of experimental trials.) 
Snapshots from three trials for each type of the ball-chains are shown 
in Figs.~\ref{fig:figure2} and \ref{fig:figure3}, respectively,
with the time interval between consecutive snapshots being $10 s$.
The difference in shapes %
between the 12 and 20 bead ball-chains is evident from the snapshots. In each figure, we present three different trials that show distinct sedimentation dynamics. In addition to the snapshots, the evolution %
is quantified by the plots of the time-dependent vertical velocity $v_y$ and bending amplitude $A/L$ for the runs shown in the snapshots.

A typical evolution of the 12-bead ball-chain is illustrated in %
Fig.~\ref{fig:figure2}. %
The ball-chain consistently shows a slightly %
distorted U-shape during all the trials. %
In Fig.~\ref{fig:figure2}(a), it seems that the ball-chain remains close to %
a plane, similar to the previous numerical %
and experimental %
studies of relatively short elastic %
fibres of a moderate stiffness %
\cite{lagomarsino2005hydrodynamic,schlagberger2005orientation,manghi2006,li2013sedimentation,saggiorato2015conformations,shojaei2015sedimentation,marchetti2018deformation, bukowicki2018different,bukowicki2019sedimenting}. 

The existence of a stable stationary U-shaped vertical configuration of an elastic filament was predicted numerically, provided that the bending stiffness is above a certain threshold value \cite{lagomarsino2005hydrodynamic,li2013sedimentation,saggiorato2015conformations,shojaei2015sedimentation,bukowicki2018different,bukowicki2019sedimenting}, and was confirmed experimentally \cite{marchetti2018deformation}. Dynamics of ball-chains have not been studied numerically so far, but it could be expected that they are similar to the dynamics of very elastic fibres.  %
Our experimental results indicate that shorter ball-chains indeed tend to a certain bent U-shape, with some small deviations.
For example, in Fig.~\ref{fig:figure2}(a), the fibre is almost planar, and this plane is slightly inclined with respect to vertical, and a small sideways drift is observed, as predicted by numerical simulations of moderately elastic filaments performed in Ref. \cite{saggiorato2015conformations}, and illustrated in the middle panel of their Fig.~1. 

We also observe behaviour that has not been reported %
in previous studies of elastic %
fibres.  In Fig.~\ref{fig:figure2}(b), the ball-chain rotates %
around the gravity direction in a screw-like fashion as it sediments. Previous studies have shown the screw-like rotation of an elastic %
fibre when under the influence of a second fibre \cite{saggiorato2015conformations}, but not in the case of a fibre sedimenting alone. The rotation is caused by small deviations from a planar vertical shape, such that the rotational-translational mobility is different from zero.

In Fig.~\ref{fig:figure2}(c), one can also see, in addition to the screw-like rotation, that the shape clearly has no left-right symmetry, and the end points of the ball-chain are not horizontally aligned. For some of the trials, like those shown in Figs.~\ref{fig:figure2}(a) and (c), the deviation from the horizontal orientation of the end-to-end vector increases with time, contrary to the expectations that a flexible relatively short filament would tend to a stable configuration with the left-right symmetry. Such an attracting symmetric configuration was found numerically for relatively stiffer elastic fibres \cite{lagomarsino2005hydrodynamic,schlagberger2005orientation,manghi2006,saggiorato2015conformations,bukowicki2018different,bukowicki2019sedimenting}. For moderate stiffness, out-of-plane shapes found in Ref. \cite{saggiorato2015conformations} also exhibit left-right symmetry.

The difference between the dynamics observed in Figs.~\ref{fig:figure2}(a), \ref{fig:figure2}(b) and \ref{fig:figure2}(c) is likely due to the difference between the initial configurations %
when the ball-chains are inserted into the fluid. However, it is worth noting that the small differences in shapes, observed in Fig.~\ref{fig:figure2}(a)-(c), result in the comparable vertical velocity of sedimentation $v_y$ and the bending amplitude $A/L$, shown in Fig.~ \ref{fig:figure2}(d) %
(the differences are not greater than 10\%). In other words, the velocity of the ball-chain exhibiting rotation and a clear lack of the left-right symmetry, with the pronounced end-to-end asymmetry, shown in Fig.~\ref{fig:figure2}(c), is very close to that of the ball-chain exhibiting only rotation and almost no end-to-end asymmetry, as seen in Fig.~\ref{fig:figure2}(b). %

The common feature of the dynamics of the shorter ball-chains is the lack of a uniquely defined stable configuration reached at a relatively short time of the evolution. Rotation and breaking of the left-right symmetry seem to be typical. 

The 20-bead ball-chains exhibit a distinctly wider range of different shapes than the 12-bead ball-chains, as seen in Fig.~\ref{fig:figure3}. We %
notice, unlike in the 12-bead runs, that the %
first observed shape of 20-bead ball-chains %
is different in each run. From the point of view of Camera 2, we see that the shape of the ball-chain as it enters the field of view is: a %
bimodal W-shape, as in Fig.~\ref{fig:figure3}(a); a %
wide U-shape, as in Fig.~\ref{fig:figure3}(b); an asymmetric, hook-shape, as in Fig.~\ref{fig:figure3}(c). The initial shapes in Fig.~\ref{fig:figure3}(a-b) %
are slightly asymmetric, %
but the initial 
shape in Fig.~\ref{fig:figure3}(c) significantly breaks the left-right symmetry. 

Fig.~\ref{fig:figure3} illustrates that ball-chain trajectories and the evolution of their shapes are sensitive to the initial configuration. The perturbed bimodal W-shapes, observed as the first snapshots, deform with time, in agreement with the numerically observed instability of the W-shape of very elastic, relatively long filaments, reported in Ref. \cite{lagomarsino2005hydrodynamic}. Hence, in Fig.~\ref{fig:figure3}(a), the initial W-shape evolves into the hook-shape before bending out of the plane and drifting away from the centre of the tank. %
On the other hand, in Fig.~\ref{fig:figure3}(c), the ball-chain is already at the hook-phase as it enters the field of view, and continues to bend out of plane and drift away from the centre of the tank. These two runs can thus be said to be similar to each other, but ``shifted in time%
” with respect to each other. 

The first shape observed in Fig.~\ref{fig:figure3}(b) resembles a deformed U-shape rather than a deformed W-shape, and it changes in time differently from Figs.~\ref{fig:figure3}(a) and (c), but 
 a bit similar to the evolution of the 12 bead ball-chain, shown in Fig.~\ref{fig:figure2}(b). However, the observed shapes are significantly non-planar and with a pronounced asymmetry, in contrast to %
 the shorter ball-chains. Moreover, they do not seem to converge to a stable planar and symmetric U-shape. This seems to be in agreement with the non-existence of a stable stationary configuration for sufficiently elastic filaments \cite{lagomarsino2005hydrodynamic,saggiorato2015conformations,bukowicki2018different}.

The stark ranges in the ball-chain shapes from the W and hook-phases to
perturbed %
rotating U-shapes indicate that the shapes are %
sensitive to the initial configuration. %
It is thus tempting to attribute this variation to the uncertainty of keeping the ball-chains horizontal and straight as they are inserted
into the tank. %
We have observed runs that, for instance, exhibit the hook-phase, but do not exhibit the significant out-of-plane bending, which can be related to only a small initial perturbation of the planar shape.  It %
might be possible that the out-of-plane bending would have eventually been observed if the tank was higher, and a small unstable perturbation had enough time to grow. The behaviour of the 20 bead ball-chains is thus 
complex, and it is difficult to classify it, %
as it was possible for the 12 bead ball-chains. %

It is remarkable that our observation of the subsequent W,  hook, and out-of-plane bending agrees very well with the %
trajectory observed in previous numerical studies of very elastic (semi-flexible) fibres at large bending amplitudes, shown in the right panel of Fig.~1 in Ref. \cite{saggiorato2015conformations}. %
Even though our experimental observations of the trajectory span only a part of the time of the numerical simulation, the agreement is striking.

The plots of the vertical sedimentation velocity $v_y$ and the bending amplitude $A$ for our experiments, shown in Fig.~\ref{fig:figure3}(a)-(c), are
presented in Fig.~\ref{fig:figure3}(d). A correlation between both parameters is visible. As $A$ increases, so too does $v_y$, and a reduction in $A$ leads to a reduction in $v_y$. The effect of the dynamic variation of the shape of the ball-chains on the vertical sedimentation velocity is evident from these snapshots, resulting in similar curves of $A$ and $v_y$
\footnote[3]{
    For some of the trials with 20-bead ball-chains, the correlation is not as evident, which is related to the more complex shapes of the ball-chains.
}.

\begin{figure*}[t!]
    \centering
    \includegraphics[width = 0.95\textwidth]{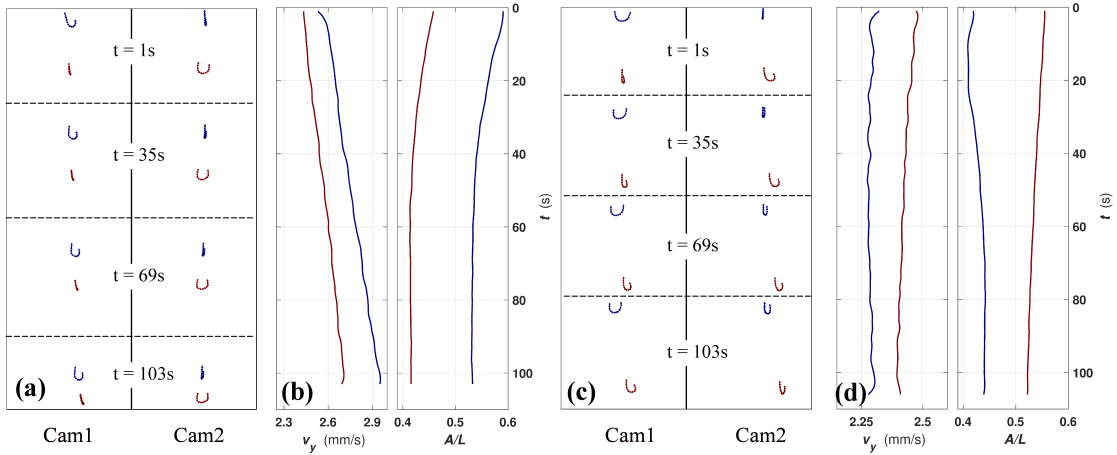}\vspace{-.3cm}
    \caption{Results of two experimental trials
    for a 
    a pair of 12-bead ball-chains %
    settling close to each other under gravity in a viscous oil, traced simultaneously by two cameras (Cam1 and Cam2) located at the same level, and with perpendicular lines of sight: %
    (a) shows snapshots of a trial in which the ball-chains come closer to each other; (c) shows snapshots of a trial in which the ball-chains separate from each other; (b) and (d) show the sedimentation velocities $v_y$ and the bending amplitudes $A/L$ of both ball-chains vs. time, plotted for the trials in (a) and (c), respectively. The blue colour represents the top ball-chain while the red colour represents the bottom ball-chain. 
    Experimental movies that correspond to the trials shown in (a) and (c) are shown in the ESI\dag{} in Videos 6a and 6c, respectively.
    }
    \label{fig:figure4}
\end{figure*}
\begin{figure*}[t!]
    \centering
    \includegraphics[width = 0.95\textwidth]{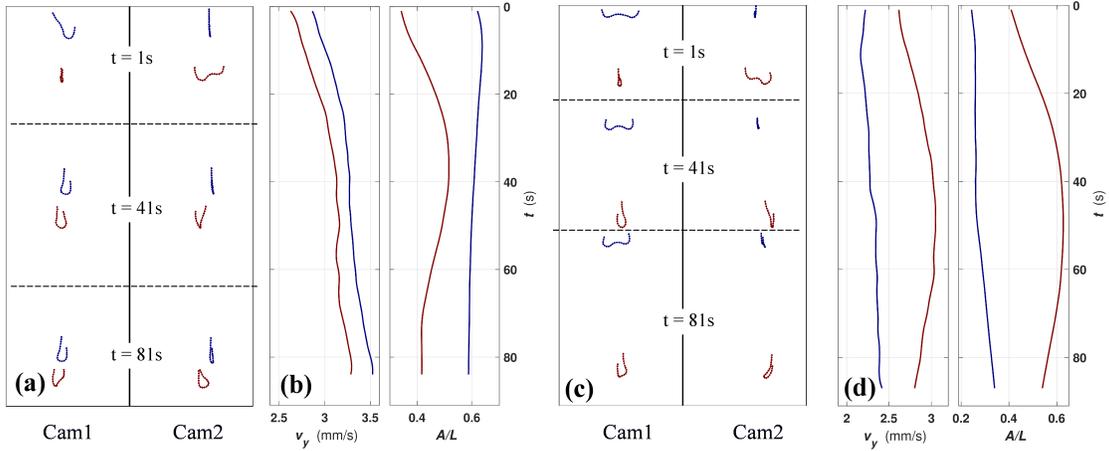}\vspace{-.3cm}
    \caption{%
    Results of two experimental trials for a pair of 20-bead ball-chains, settling close to each other under gravity in a viscous oil, traced simultaneously by two cameras (Cam1 and Cam2) located at the same level, and with perpendicular lines of sight: %
    (a) shows snapshots of a trial in which the ball-chains come closer to each other; (c) shows snapshots of a trial in which the ball-chains separate from each other; (b) and (d) show the sedimentation velocity $v_y$ and the bending amplitude $A/L$ vs. time, plotted for the trials in (a) and (c), respectively. The blue colour represents the top ball-chain while the red colour represents the bottom ball-chain.
    Experimental movies that correspond to the trials shown in (a) and (c) are shown in the ESI\dag{} in Videos 7a and 7c, respectively.
    }
    \label{fig:figure5}
\end{figure*}

\subsection{Sedimentation of Two Interacting Ball-chains} \label{two_ballchains}

\subsubsection{Experimental Observations}

In order to study the hydrodynamic interactions of two ball-chains as they settle under gravity in a viscous fluid, %
we choose an initial configuration as follows: the ball-chains are inserted horizontally one above the other, such that their centres are aligned vertically, with the bottom ball-chain being in the focal plane of Camera 1 (and perpendicular to Camera 2) and the top ball-chain being in the focal plane of Camera 2 (and perpendicular to Camera 1). The ball-chains are inserted manually at the centre of the tank. The motion of both ball-chains is recorded using the camera arrangement shown in Fig.~\ref{fig:figure1}. We performed a %
number of trials, i.e., 13 trials for the 12-bead ball-chains and 23 trials for the 20-bead ball-chains, to study time-dependent %
configurations of the sedimenting ball-chains. 

We present the typical behaviour of a pair of ball-chains in Figs.~\ref{fig:figure4} and \ref{fig:figure5} for the 12-bead and 20-bead ball-chains, respectively. In both Figs.~\ref{fig:figure4} and \ref{fig:figure5}, we present two individual trials, shown in panels (a)-(b) and (c)-(d), respectively. In addition to the snapshots visible in panels (a) and (c), we also show in panels (b) and (d) the corresponding bending amplitudes $A$ and the sedimentation velocities $v_y$ of the top and bottom fibres.

In Fig.~\ref{fig:figure4}(a), we observe that the 12-bead ball-chains approach  each other as they sediment (the term ``attraction” is used to describe such a behaviour). 
It is shown in Fig.~\ref{fig:figure4}(b), that the top ball-chain (coloured blue) exhibits a larger bending amplitude, as well as a higher sedimentation velocity than the bottom ball-chain (coloured red). In other words, the top ball-chain bends 
in such a way that its mobility is larger 
than the mobility of the bottom ball-chain. %

Previous studies of elastic and semiflexible fibres have also demonstrated the attraction between two fibres of low/moderate elasticity that are one above the other \cite{llopis2007sedimentation, saggiorato2015conformations}. In our experiments with the 12-bead ball-chains, we have consistently observed that the attraction of the ball-chains is %
correlated with the top ball-chain having a larger bending amplitude than the bottom one. This finding is in agreement with the results of numerical simulations, shown in Fig.~5 in Ref. \cite{saggiorato2015conformations}, where  $A_{\text{top}} > A_{\text{bot}}$. Attraction of elastic fibres settling one above the other was also found in numerical simulations presented in Ref. \cite{llopis2007sedimentation} 
\footnote[4]{
    Numerically determined shapes of elastic fibres sedimenting within the same vertical plane, presented in  Fig.~2 in Ref. \cite{llopis2007sedimentation}, are different from those shown in Fig.~5 in Ref. \cite{saggiorato2015conformations}, probably owing to the use of a different, simplified theoretical model.
}.

We also present, to the best of our knowledge, the first demonstration of a vertical separation of ball-chains that are one above the other. %
This behaviour has been observed in our experiments, as clearly visible in the snapshots in Fig.~\ref{fig:figure4}(c). The bottom ball-chain (coloured red) moves away from the top ball-chain (coloured blue) as both ball-chains sediment. %
The plots of sedimentation velocities $v_y$ and bending amplitudes $A$ of both fibres, presented in Fig.~\ref{fig:figure4}(d), corroborate the separation of the ball-chains, with the bottom ball-chain having a larger bending amplitude, and, consequently, a larger sedimentation velocity than the top ball-chain. 

In our experiments, the shapes of the ball-chains typically do not have the right-left symmetry, and the non-horizontal end-to-end line of the ball-chains seems to increase significantly their speed. On the other hand, deviations from the horizontal location of the ball-chain arms seem to be random, and for the top ball-chain, they can be both smaller and larger than for the bottom one. We have not observed a tendency to approach with time symmetric or even planar shapes, nor any specific relative orientation of the ball-chains. 

The 20-bead ball-chains also exhibit the two distinct behaviours described for the 12-bead ball-chains: vertical attraction or repulsion. %
Fig.~\ref{fig:figure5}(a) shows snapshots from a trial where %
the ball-chains approach each other, while Fig.~\ref{fig:figure5}(c) shows snapshots from another trial where the ball-chains move away from each other. In most runs, we observe a direct correlation between the bending amplitude $A$ and the sedimentation velocity $v_y$ - the larger the bending amplitude, the faster %
the sedimentation velocity (see %
Fig.~\ref{fig:figure5}(b) and (d)), but there are exceptions%
. %
If the top ball-chain bends more than the bottom ball-chain, the ball-chains come together. If the bottom ball-chain bends more than the top ball-chain, a separation is observed. 

\subsubsection{How Much Hydrodynamic Interactions between Two Ball-Chains Change the Isolated Ball-Chain Dynamics?}

Previous studies on the dynamics of two very %
elastic filaments in a top-down initial orientation suggest that at large values of $B$%
, the dynamics of the filaments are dominated by the behaviour of a single filament, and not by hydrodynamic interactions between the two filaments, if the distance between the filaments is not small  \cite{saggiorato2015conformations}. The shapes exhibited by the separating 20-bead ball-chains that are relatively far from each other in an experimental trial shown in Fig.~\ref{fig:figure5}(c) seem to agree with this assessment, since the sequence of the ball-chain shapes is close to the different stages of their individual sedimentation process - for instance, both ball-chains in the first frame of  Fig.~\ref{fig:figure5}(c) are in the bimodal W-phase, but the bottom ball-chain is already beginning to enter the hook-phase. Such a time shift could be easily understood since the upper ball chain was inserted into the fluid later than the lower one. 

\begin{figure*}[t!]
 \centering
 \begin{tabular}[b]{c}
  \includegraphics[width=0.45\textwidth]{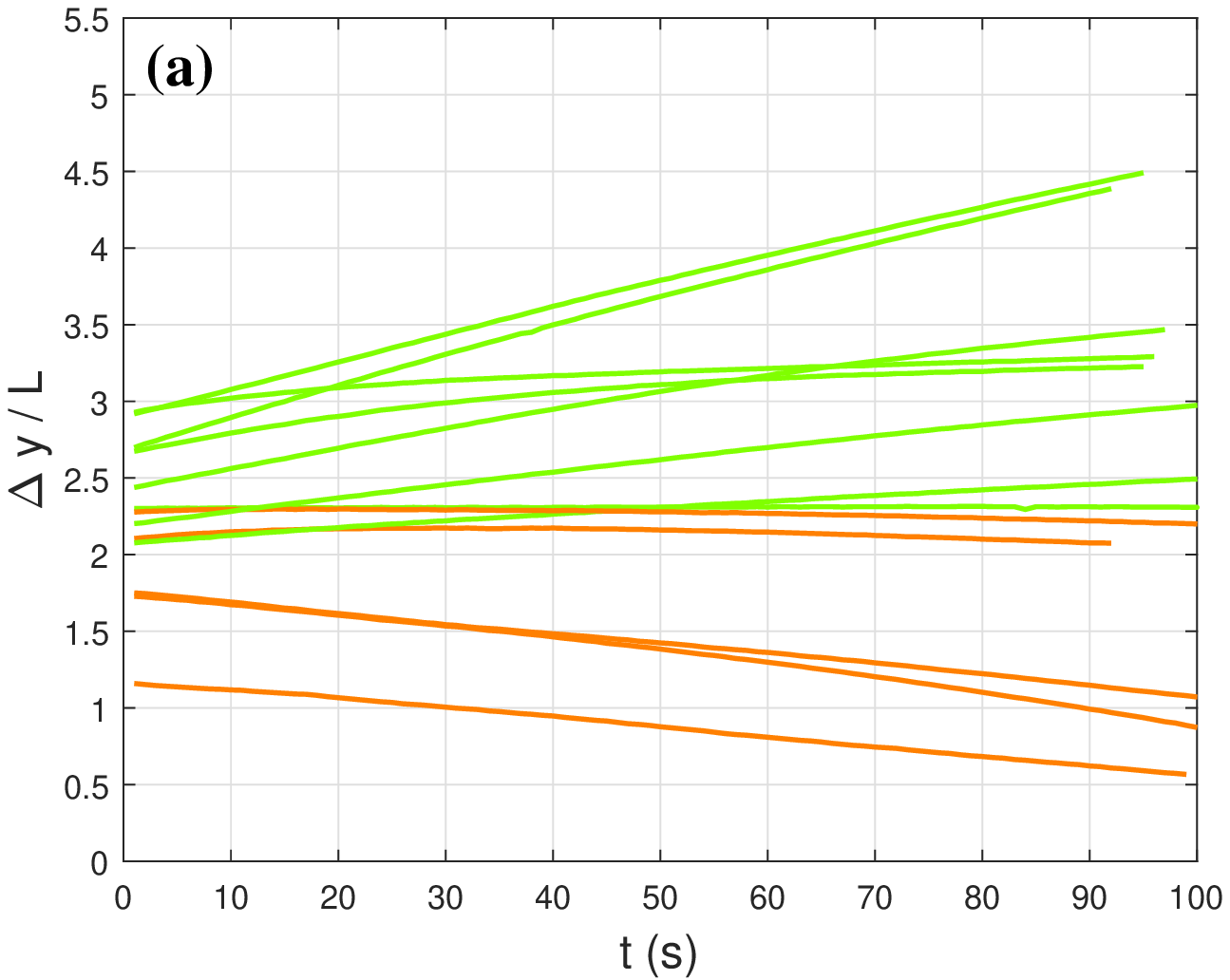}
 \end{tabular}
 \begin{tabular}[b]{c}
  \includegraphics[width=0.45\textwidth]{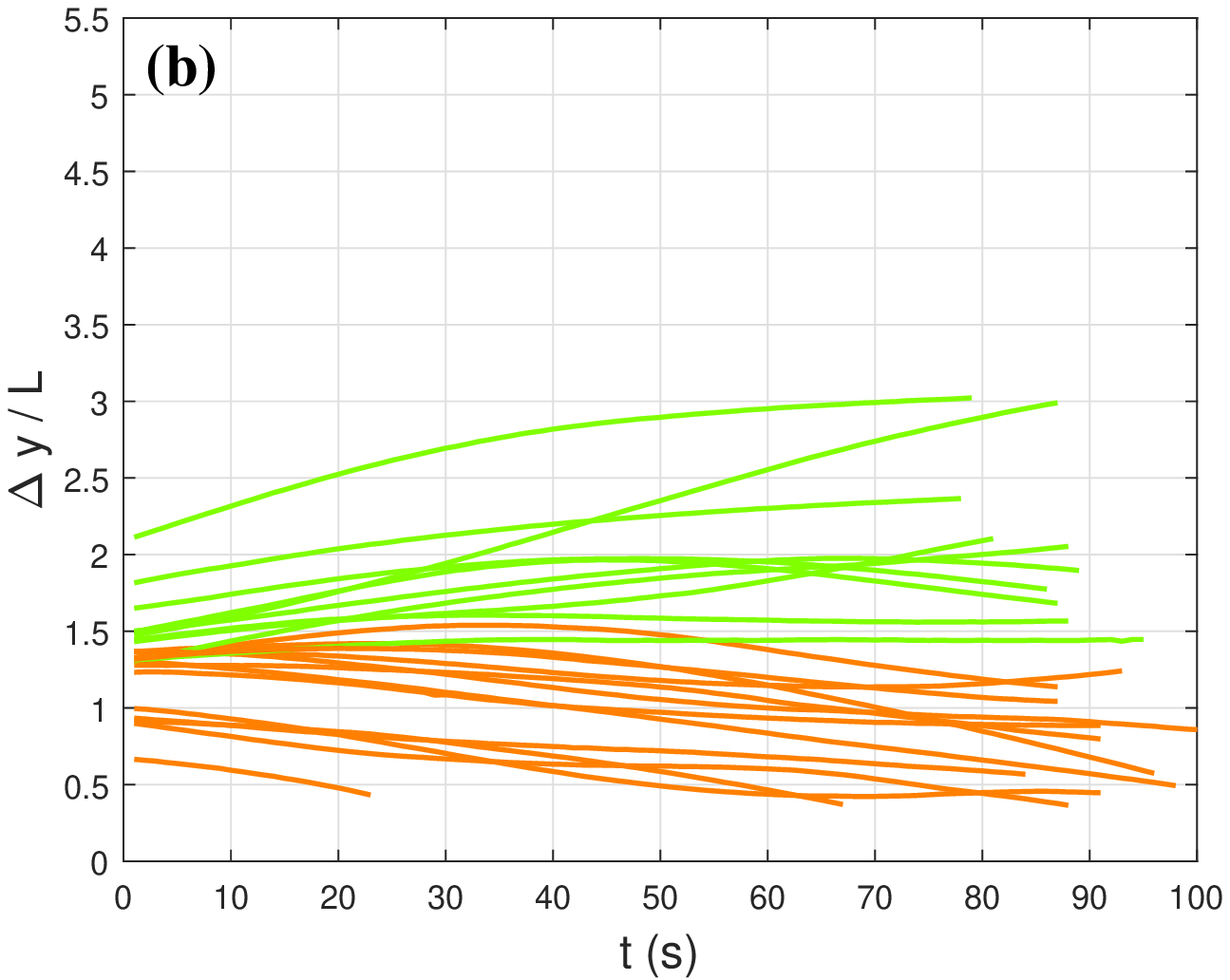}
 \end{tabular}\vspace{-0.3cm}
 \caption{%
 The vertical distance between the ball-chains, $\Delta y$,  normalised with ball-chain length $L$, as a function of time for: (a) 12 beads, (b) 20 beads. The orange curves represent ball-chains that come closer to each other and the green ones ball-chains that move away from each other.}
 \label{fig:figure7}
\end{figure*}
\begin{figure*}[t!]
 \centering
 \begin{tabular}[b]{c}
  \includegraphics[width=0.45\textwidth]{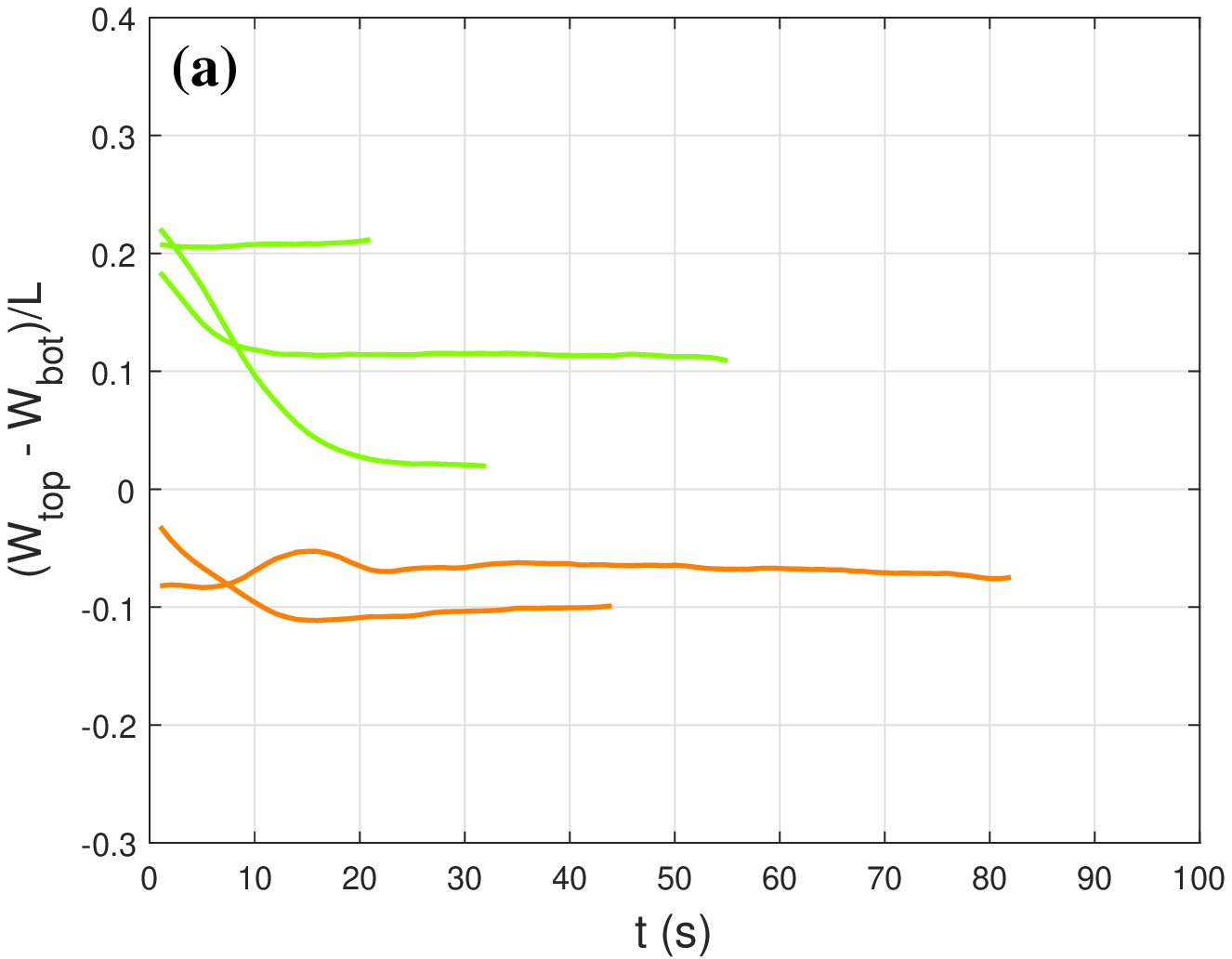}
 \end{tabular}
 \begin{tabular}[b]{c}
  \includegraphics[width=0.45\textwidth]{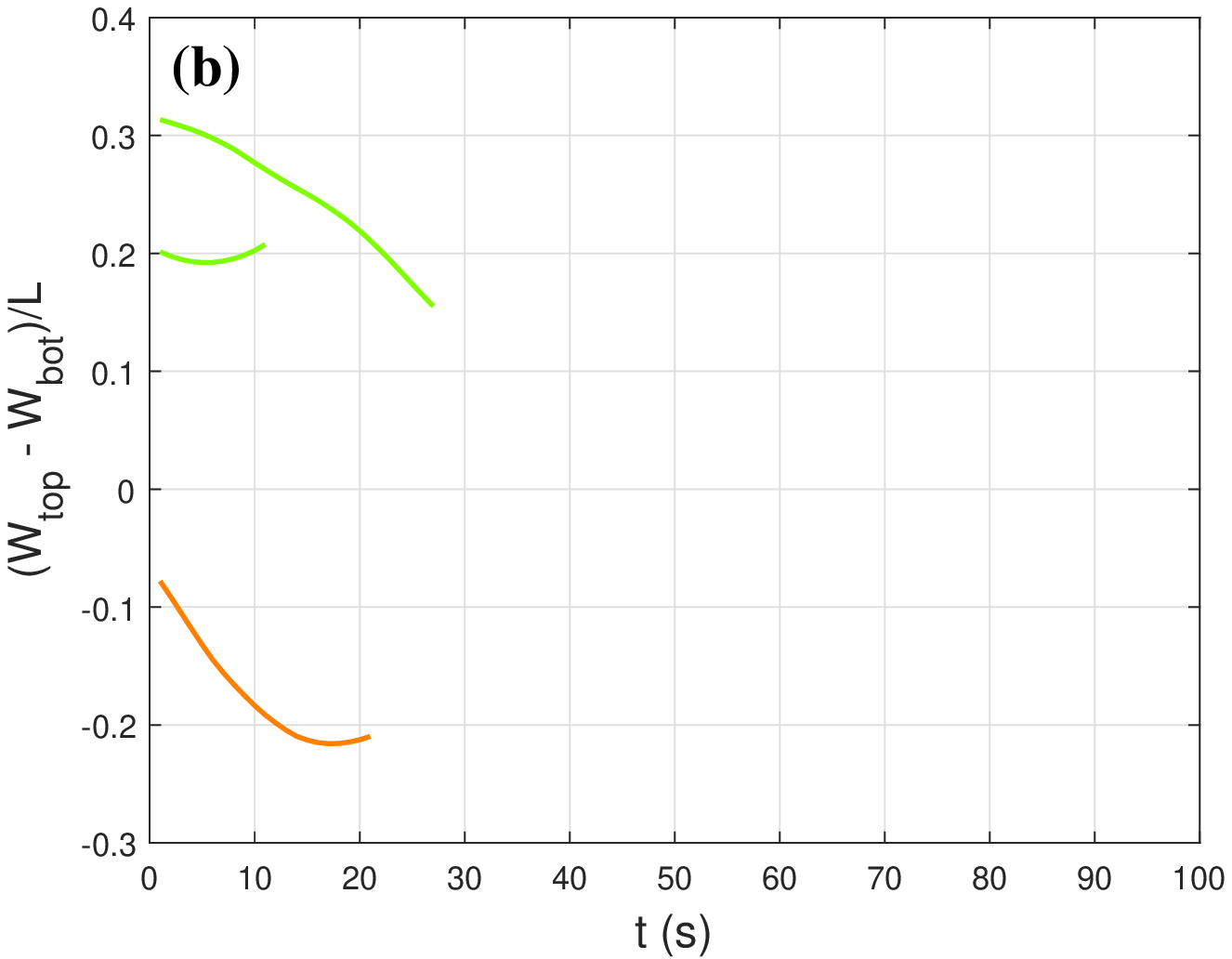}
 \end{tabular}\vspace{-0.3cm}
 \caption{Difference  $(W_\text{{top}} - W_\text{{bot}})/L$ of the top and bottom ball-chain widths %
 as a function of time, for: (a) 12 beads, (b) 20 beads. The orange curves represent ball-chains that approach each other, and the green ones represent ball-chains that move away from each other.}%
 \label{fig:figure8}
\end{figure*}
The velocity of a hook-shaped ball-chain is larger than that of a W-shaped ball-chain, as discussed earlier and shown in Fig.~\ref{fig:figure3} of Sec. \ref{single_ballchain}. Furthermore, %
by closely observing the values of the sedimentation velocity $v_y$ of both the top and bottom ball-chains (Fig.~\ref{fig:figure5} (d)), we see that the separating ball-chains exhibit a similar velocity range to that of an isolated ball-chain (Fig.~\ref{fig:figure3} (d)). %
It is thus very likely that the separation of the 20-bead ball-chains observed in the current experiments is merely a result of the bottom ball-chain being further in its individual evolution than the top ball-chain. We also point out that, while it is possible to explain the separation of the 20-bead ball-chains based on their sequences of shapes, %
the limited range of shapes %
exhibited by the 12-bead ball-chains makes it difficult to explain the reason of their separation. 

For the top and bottom ball-chains relatively close to each other, it is expected that their mutual hydrodynamic interactions would cause them to approach each other \cite{saggiorato2015conformations}.
We have seen clear evidence of an attracting behaviour of the hydrodynamic interactions in a few trials where the top ball-chain, which is not directly above the bottom ball-chain but still coming closer to the bottom ball-chain changes its trajectory and turns towards the bottom ball-chain once it is close enough (see Videos 6a and 7a in the ESI\dag{} 
). %
Direct evidence of hydrodynamic interactions can be found in Figs.~\ref{fig:figure4}(b) and \ref{fig:figure5}(b) - we observe that both ball-chains sediment at a larger velocity when compared to the sedimentation velocity of an isolated ball-chain,  shown in Figs.~\ref{fig:figure2}(d) and \ref{fig:figure3}(d). It is well known %
that a particle  %
sedimenting close to other particles
moves %
faster than in the absence of any other particles. %

To further illustrate the relevance of hydrodynamic interactions at small distances, we present in Fig.~\ref{fig:figure7} the %
time-dependence of vertical distance $\Delta y$ between the two ball-chains as they traverse the height of the tank, for all the experimental runs performed (see Fig.~\ref{fig:figure1a} for the definition of $\Delta y$). 
We adopt the following colour code in the plots: The runs with a greater vertical separation distance between the ball-chains at the end of the run than at the beginning are assigned the colour green (moving apart, separating), whereas the runs with a smaller vertical separation distance at the end of the run than at the beginning are assigned the colour orange (moving closer, attracting). 
In other words, the green curves correspond to runs in which $\Delta y_{t=end} > \Delta y_{t=0}$ and the orange curves correspond to runs in which $\Delta y_{t=end} < \Delta y_{t=0}$. 

It is clear from the plots in Fig.~\ref{fig:figure7} that the orange curves correspond to two ball-chains that are closer to each other  at the beginning of the run.
We observe this tendency for both the 12-bead as well as the 20-bead runs. 
It is known that hydrodynamic interactions between two flexible objects (or groups of particles), one above the other, cause the lower object to become wider and with a smaller vertical dimension \cite{ekiel2014,saggiorato2015conformations}. %
This mechanism can cause the lower ball-chain to move slower than the upper one. 
It is clear that %
the closer the ball-chains are to each other, the stronger they interact hydrodynamically. %

On the other hand, the green curves in Fig.~\ref{fig:figure7} correspond to two ball-chains that are further from each other at the beginning of the run. Therefore, the influence of the %
isolated ball-chain dynamics on their evolution is more pronounced. As discussed before, the time shift between the moments of the release of both 20-bead ball-chains might explain why they move away from each other in the early stage of the runs. Indeed, as shown in Fig.~\ref{fig:figure3}(d), the velocity of an isolated ball-chain increases with time in the early stage of its evolution. However, later it reaches a maximum and then decreases, which might be responsible for the non-monotonicity of some green curves in 
Fig.~\ref{fig:figure7}(b), showing that the ball-chains initially move away from each other before eventually coming closer together. Important are also
dynamic changes of shapes for the 20-bead ball-chains. %
The 12-bead ball-chains do not change their shapes so dynamically, %
which seems to be related to the almost monotonic curves in Fig.~\ref{fig:figure7}(a). 

It might be interesting to compare the widths of the top and bottom ball-chains, $W_{\text{top}}$ and $W_\text{{bot}}$, as defined in Fig.~\ref{fig:figure1a}(b). 
The ball-chains in the current experiments exhibit significant out-of-plane motion, such as rotation and bending. It is thus challenging to determine %
the width of the ball-chains %
since in general, they are %
not in a plane that would coincide with %
either plane of the camera views. However, there exist a few runs in which both ball-chains do not exhibit significant rotation or out-of-plane bending. We measure the ball-chain width $W$ of such runs, %
but we only consider frames in which both top and bottom ball-chains are planar, and their planes coincide with the perpendicular planes of the camera view.  %
In other words, the top ball-chain  needs  to stay within the plane of view of %
Camera 1 and perpendicular to the plane of view of Camera 2. At the same time, the bottom ball-chain needs  to stay within the plane of view of %
Camera 2 and perpendicular to the plane of view of Camera 1. %
Our %
analysis is thus reduced to only a few runs that satisfy these conditions. 

In Fig.~\ref{fig:figure8}, we %
plot the difference in widths of the top and bottom ball-chains, $W_\text{{top}} - W_\text{{bot}}$, %
normalised with the ball-chain length $L$, as a function of the time $t$. %
It is evident from the plot that $W_\text{{top}} > W_\text{{bot}}$
for runs in which the ball-chains move away from each other, %
and  $W_\text{{top}} < W_\text{{bot}}$
for the runs in which the ball-chains approach each other. %

The small number of curves shown in Fig.~\ref{fig:figure8} is related to a generic feature of the dynamics observed in our experiments: the ball chains tend to rotate and bend out of plane.  
There are only 5 runs (out of 13) for 12-bead ball-chains, and only
3 runs (out of 23) for 20-bead ball-chains, such that both ball chain shapes remain in a single plane, and also only for a limited period of time at the early stage of the evolution, as shown in Fig.~\ref{fig:figure8}. 
Probability that the ball-chains remain in their original planes is very small.
Our experiments indicate that the out-of-plane motion of very flexible elongated objects is ubiquitous, and that their purely planar motion %
would seldom be observed in practical situations. 

\section{Numerical Simulations}
In the experiments, the ball-chains are fully flexible until a limit bending angle is reached. They are %
assumed to move and deform in a similar way %
to very elastic fibres. To confirm this, we compare the experiments with numerical simulations of elastic fibres. Taking into account the sensitivity of the experimental results to the initial conditions, we focus our numerical analysis on a single sedimenting very elastic fibre.
The goal is to verify whether we can theoretically reproduce basic features of the dynamics observed in our experiments, i.e., a non-stationary fibre shape
which may exhibit rotation, out-of-plane deformation and 
which may have 
a non-horizontal end-to-end vector connecting the first and the last beads.

\subsection{
Theoretical Description %
}

In order to model dynamics of a single fibre settling under gravity in a viscous fluid at a Reynolds number much smaller than unity, we employ the bead-chain model in which the fibre is represented by $N$ identical spherical beads of a diameter $d$ \cite{gauger_numerical_2006}. Centres of %
consecutive %
beads are connected by springs. %
The distance between the centres of the consecutive beads (the bond length) at the elastic equilibrium %
 is $l_0$, chosen to be very close to the bead diameter, $l_0=1.01 d$. Thus the %
length of the fibre at the elastic equilibrium is $L_0=(N-1) \cdot l_0 + d$.

To represent almost inextensible elastic interactions between the consecutive beads $i$ and $j= i+1$, with $i=1,...,N-1$,  %
we employ finitely extensible nonlinear elastic (FENE) potential energy %
that has the following form \cite{warner_kinetic_1972, bird_dynamics_1977}:
\begin{equation}
\label{eq:FENE}
U^{FENE} = - \frac{1}{2} k (l_{0}-d)^{2} \sum_{i=1}^{N-1}\ln\left[ {1 - \left(\frac{r_{i,i+1}-l_{0}}{l_{0}-d} \right)^{2}} \right]. %
\end{equation}
Here, 
  $r_{i,j}=|\bm{r}_i-\bm{r}_j|$ is the distance between the centres of beads $i$ and $j$ (in particular, $r_{i,i+1}$ is the %
  length of bond $i$), %
  with
  $\bm{r}_{i}$ being a time-dependent position of bead $i$,
  and
  $k$ is the spring constant.

With the FENE potential energy defined in Eq.~\ref{eq:FENE}, %
the distance between surfaces of the consecutive beads, initially equal to $0.01d$, stays very small during settling, and does not exceed $0.02 d$. For such a %
small gap, lubrication interactions with the fluid suppress %
spurious rotations of the beads \cite{slowicka2015}. 

At the elastic equilibrium, the fibre is straight. 
During its settling, it resists deformations by bending
forces between triplets of the consecutive beads. %
We employ the harmonic bending potential energy \cite{storm_theory_2003, li_viscoelasticity_2012, bukowicki2018different, gruziel_periodic_2018, gruziel2019stokesian}: %
\begin{equation}
\label{eq:BEND}
   U^{b} =  \frac{\mathcal{A}}{2 l_{0}}  \sum_{i=2}^{N-1} %
   \beta_{i} %
   ^{2}.
\end{equation}
Here, 
 $\beta_{i}$ is the bending angle of a triplet of consecutive beads $i-1,\;i,\;i+1$, defined in Eq.~\ref{eq:beta} and shown in Fig.~\ref{fig:angle_beta}, 
and
 $\mathcal{A}$ is bending stiffness.
Based on the model of an elastic cylinder of diameter $d$ \cite{landau_1986}, the expressions
\begin{equation}
  \begin{cases}
   k = E_{Y} \pi d^2 / \left( 4 l_{0} \right) \\
   \mathcal{A} = E_{Y} \pi d^4 /  64 
  \end{cases}
\end{equation}
allow identification of the spring constant $k$ and the bending stiffness $\mathcal{A}$ if the Young's modulus $E_{Y}$ is known \cite{bukowicki2018different}.

The total elastic potential energy of the fibre %
has the form,
\begin{equation}
U = %
U^{FENE}
+ %
U^b.
\end{equation}
The %
elastic force on a bead $i$ is
\begin{equation}
\bm{F}_i^e=- \frac{\partial}{\partial \bm{r}_i} U.
\end{equation}

In addition, each bead is subject to the same constant external gravitational force, %
$\bm{F}^g = - m g \cdot \mathbf{ \hat{y} }$, where $g$ is the gravitational acceleration, $ \mathbf{ \hat{y} }$ is the unit vector along $y$ axis,  and $m$ is the bead mass corrected for buoyancy. The total external force on bead $i$ reads,
\begin{equation}
\bm{F}_i=\bm{F}_i^e + \bm{F}^g.
\end{equation}

Dimensionless elasto-gravitational number $B$ estimates the ratio of gravitational to bending forces %
in the following way \cite{llopis2007sedimentation,lagomarsino2005hydrodynamic,saggiorato2015conformations,marchetti2018deformation,bukowicki2018different,bukowicki2019sedimenting}:
\begin{equation}
    B = L_{0}^2 Nmg %
    / \mathcal{A},
\end{equation}
with larger values of $B$ corresponding to fibres that are more flexible. %

\begin{figure*}[t!]
    \centering
    \includegraphics[width = 0.95\textwidth]{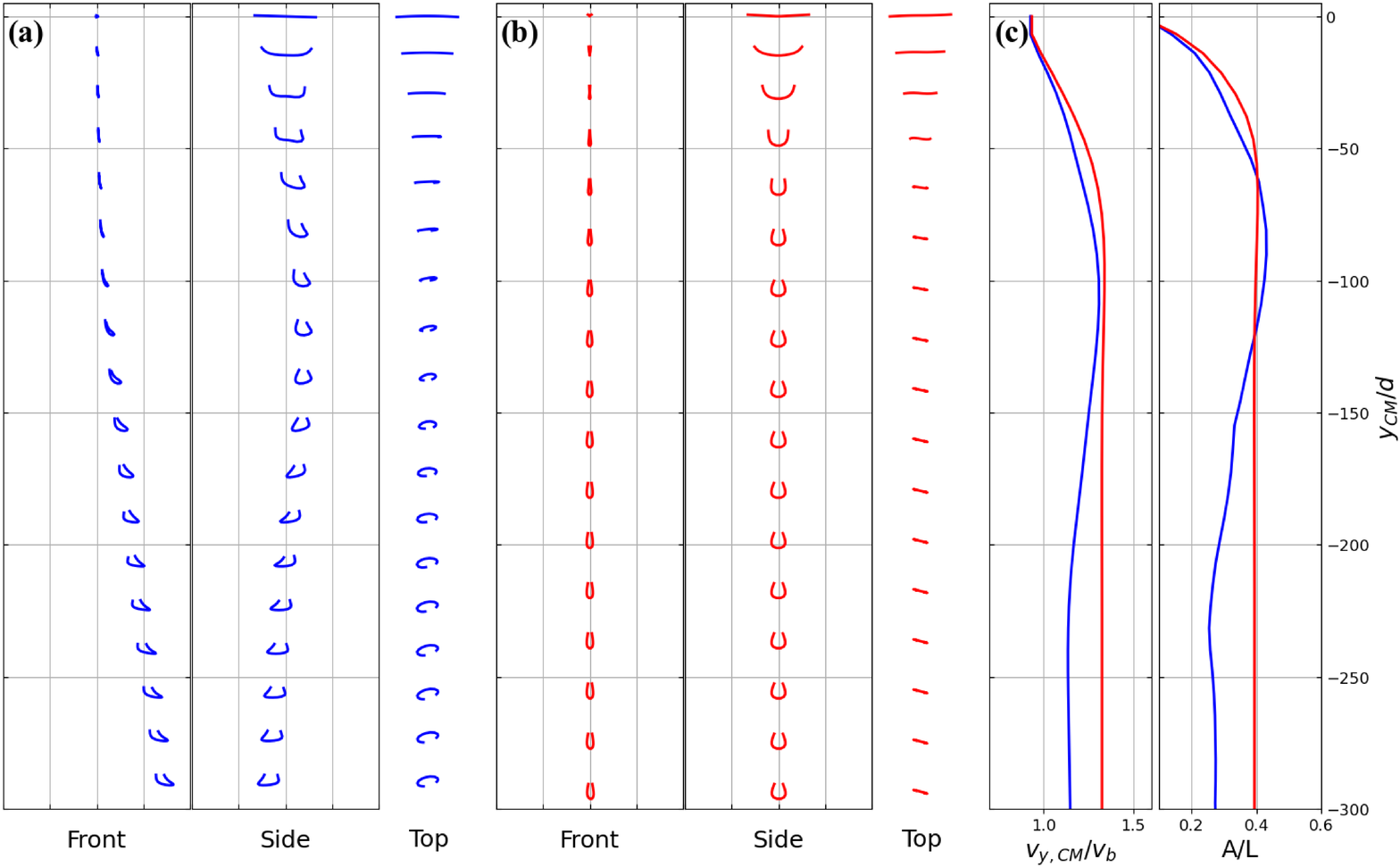}
    \caption{
    Simulation results for a single elastic fibre made of 14 beads  settling under gravity. (a), (b) 
    Snapshots taken %
    at the %
    time intervals 
    $\Delta t = 14.5 \tau_b$ 
    for %
    (a) $B=8500$ and initially C-shape, and
    (b) $B=7000$ and initially symmetric propeller shape.
    (c) The sedimentation velocity of the centre of mass $v_{y,CM}/v_b$ and the bending amplitude $A/L$ vs. vertical component of the center-of-mass position $y_{CM}/d$, %
    plotted with blue and red lines for the cases shown in (a) and (b), respectively.
    }
    \label{fig:figure_simulations_14}
\end{figure*}

As the Reynolds number in our experiments is %
very small, $Re \ll 1$, %
we assume that the fluid flow satisfies the Stokes equations. We use the multipole expansion of solutions to the Stokes equations \cite{kim,cichocki1994}, and assume stick boundary conditions on the surfaces of the beads. Then, we introduce the lubrication correction to speed up the convergence of the multipole expansion \cite{durlofsky,cichocki_lubrication_1999}. We obtain the following  set of the first-order ODEs for the time-dependent positions $\bm{r}_i$ of the bead centers, $i=1,...,N$, 
\begin{equation}
\dot{\bm{r}}_{i} = \sum_{j=1}^{N} \bm{\mu}_{ij} \cdot \bm{F}_j, %
\end{equation}
with the mobility matrices $\bm{\mu}_{ij}$ that depend on the positions of all the beads, and are evaluated from the multipole expansion by the precise numerical codes Hydromultipole \cite{cichocki_lubrication_1999,ekiel-jezewska_precise_2009}.

The numerical codes Hydromultipole are capable of evaluating the mobility matrices in the presence of interfaces \cite{cichocki_friction_2000,cichocki2004motion,cichocki2007hydrodynamic,bhattacharya_many-particle_2005,bhattacharya2005hydrodynamic,blawzdziewicz2010motion}. However, here we applied the model of an infinite fluid, since during the experimental observations the fibres were relatively far away from the container walls and the free surface.

Here we have used dimensionless variables, using as the length and time units 
$d$ and $\tau_b = \frac{\pi \eta d^2}{mg}$, respectively, where $\eta$ is the fluid dynamic viscosity. Therefore, velocity unit is
$v_b = \frac{d}{\tau_b} = \frac{mg}{\pi \eta d}$. %

\subsection{Results: Sedimentation of a Single Elastic Fibre}

\begin{figure*}[t!]
    \centering
    \includegraphics[width = 0.95\textwidth]{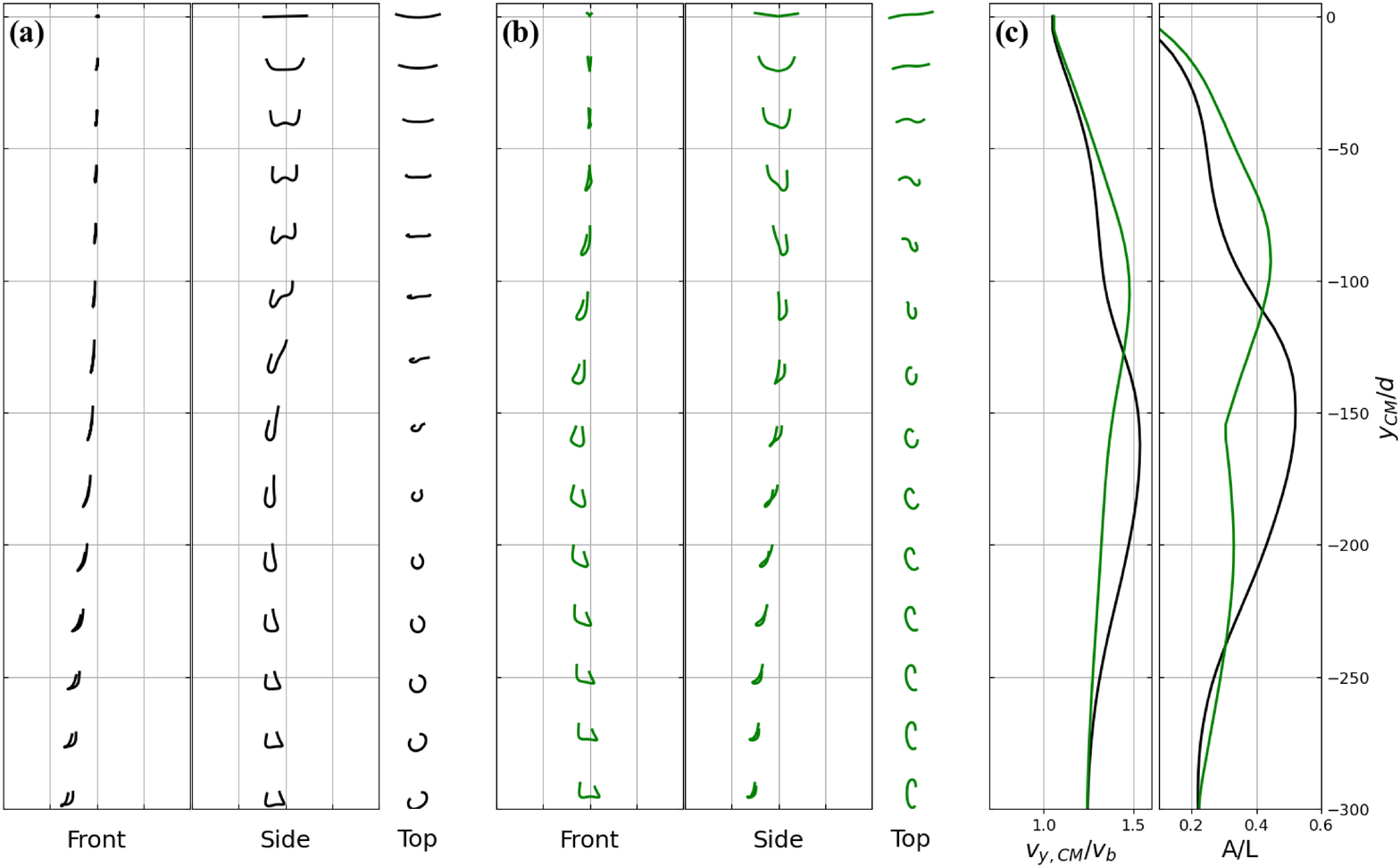}
    \caption{Simulation results for a single elastic fibre made of 24 beads  settling under gravity. (a), (b) 
    Snapshots taken %
    at the %
    time intervals 
    $\Delta t = 17 \tau_b$ 
    for %
    (a) $B=8500$ and initially C-shape, and
    (b) $B=8500$ and initially asymmetric propeller shape.
    (c) The sedimentation velocity of the centre of mass $v_{y,CM}/v_b$ and the bending amplitude $A/L$ vs. vertical component of the center-of-mass position $y_{CM}/d$, %
    plotted with blue and red lines for the cases shown in (a) and (b), respectively.
    }
    \label{fig:figure_simulations_24}
\end{figure*}

\begin{figure}[ht!]
    \centering
    \includegraphics[width=6.7cm]{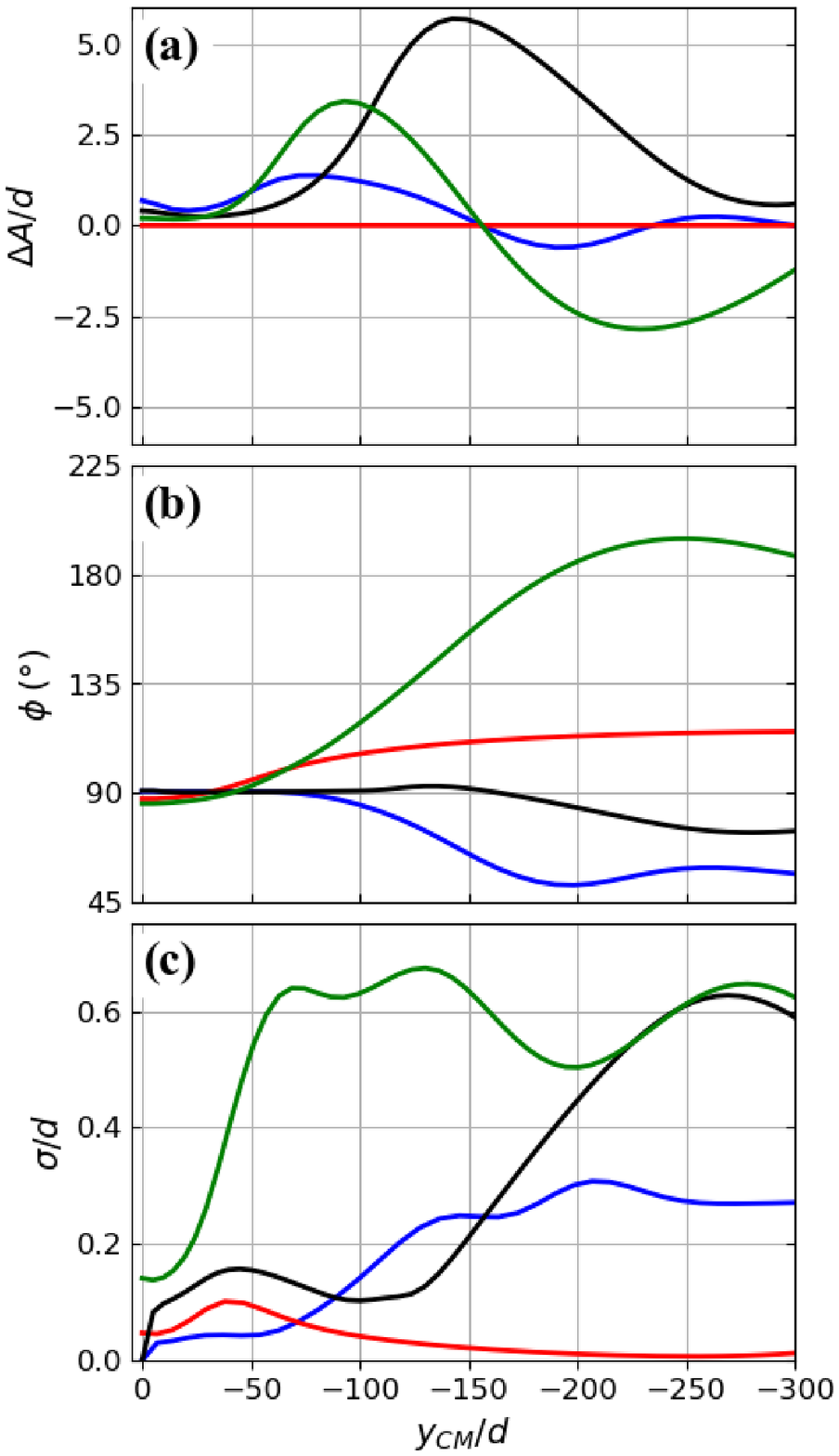}
    \vspace{-.5cm}
    \caption{
    (a) Difference $\Delta A/d$ between vertical positions of the fibre ends, %
    (b) azimuthal angle $\phi$ %
    and
    (c) %
    non-flatness $\sigma/d$, plotted  %
    as functions of vertical component of the center-of-mass position  $y_{CM}/d$ %
    for the numerical trials shown in Figs.~\ref{fig:figure_simulations_14} and 
    \ref{fig:figure_simulations_24}, with the same meaning of the colours.
    }
    \label{fig:figure_simulations_auxplot}
\end{figure}

Numerical simulations were performed of the dynamics of a single fibre having different numbers $N$ of the beads. Here we present the results for either $N=14$ (shorter fibre) or $N=24$ (longer fibre). The aspect ratios of these fibres were chosen to be similar to the aspect ratios of the ball-chains in our experiments. 

Very flexible fibres were assumed in this study, with the elasto-gravitational number 
in the range $4,000<B<10,000$.
The choice of a very large $B$ was obvious to model %
very flexible ball-chains in the experiment. Specific choice of values was guided by the requirement that 
the elasto-gravitational number is large enough to allow for %
excitation of higher bending modes, with out-of-plane deformations, and lack of stability of a U-shaped configuration, reported in \cite{lagomarsino2005hydrodynamic,saggiorato2015conformations}. This condition was needed  to match our experimental observations.  

Due to the experimental conditions, we are interested in relatively short times of the evolution that allow to study only the initial stage of the settling dynamics. Before the beginning of the observations, while settling on the distance of 10 cm, i.e., 67$d$, the ball-chains already reach U-shaped almost planar configurations. For very flexible elastic fibres characteristic times to bend are also very short \cite{lagomarsino2005hydrodynamic}. The height of the observed part of our %
glass tank is about 300 mm, i.e., about 200$d$, %
and therefore simulations should be able to 
cover about $300 d$ in the vertical direction. This distance is reached at about $255 \tau_{b}$ in case of a shorter fibre with $N=14$ beads, which is slower than about $235 \tau_{b}$ required for a longer one with $N=24$ beads.
Nevertheless, in either case, this time is usually smaller than the time required to %
reach 
one of stationary or stable periodic  %
modes of sedimentation. Moreover, time needed to destabilize U-shaped or W-shaped configurations is very sensitive to its out-of-plane perturbations. %

Therefore, it is essential to choose proper initial conditions in the simulations, allowing for a fast destabilisation and triggering out-of-plane dynamics as early as possible  rather than matching 
precisely the initial configurations observed in the experiments. It is known that initially straight fibre, horizontal or inclined, does not destabilize fast even if a small random perturbation is added \cite{lagomarsino2005hydrodynamic,saggiorato2015conformations}, as discussed in Appendix \ref{straight}.
Taking it  into account, we searched for initial configurations close to elastic equilibrium but not in a vertical plane. We used such configurations: an inclined planar C shape with a non-horizontal end-to-end vector (in short, C-shape), and a propeller shape, with horizontal and non-horizontal end-to-end vector (in short, symmetric and asymmetric propeller shapes, respectively). These three types of the initial configurations are specified in details in Appendix \ref{aA}. It is essential that the initial configurations are only slightly disturbed in comparison to a straight fibre in elastic equilibrium, as shown in the first snapshots in Figs. \ref{fig:figure_simulations_14}a,b and Fig. \ref{fig:figure_simulations_24}a,b.

The elastic fibres made of 14 beads, shown in Fig. \ref{fig:figure_simulations_14}, %
rotate. It is a generic feature seen in our numerical simulations, in agreement with the experimental observations presented in Figs. \ref{fig:figure2}b, \ref{fig:figure2}c and the corresponding Videos 4b, 4c. In the simulations, initial shapes with a slightly non-horizontal end-to-end position seem to have one arm higher than the other for a long time, also in agreement with the experiments, as illustrated in Fig. \ref{fig:figure2}. In addition, in the simulations such shapes %
significantly deform out of a vertical plane, as shown in Fig. \ref{fig:figure_simulations_14}a. 

Simulations of longer elastic fibres, made of 24 beads, also demonstrate an evolution pattern similar to that seen in the experiments (compare Fig. \ref{fig:figure_simulations_24} to Fig. \ref{fig:figure3}). The fibres form a W shape, then tilt it and form a hook, deform out of a vertical plane while decreasing the difference between vertical coordinates of the ends. 

It is interesting to point out that the formation of the W shape depends not only on the value of $B$\cite{saggiorato2015conformations}, but also on the fibre aspect ratio $N$. It does not take place for shorter fibres made up of 14 beads, while it is present for longer fibres made up of 24 beads, with the same value of the elasto-gravitational number $B$. 

In connection with the experimental findings, it is worth investigating in the simulations if the initially imposed small left-right asymmetry %
of the fibre shape will decrease or increase with time, or oscillate or stabilize. To see it, we evaluate for each time the difference $\Delta A=y_{N}-y_{1}$ between vertical coordinate, $y_N$ and $y_1$, of the positions of the centres of the last and the first bead of the fibre. Then, $\Delta A/d$ is plotted in Fig. \ref{fig:figure_simulations_auxplot} as a function of the dimensionless instantaneous vertical coordinate $y_{CM}/d$ of the fibre centre of mass, for the numerical trials shown in Figs.~\ref{fig:figure_simulations_14} and \ref{fig:figure_simulations_24}, %
with the same meaning of the colours. It is clear that the initially small (but non-zero) value of $\Delta A=y_{N}-y_{1}$ increases when the fibre settles down and then oscillates for quite a long time. Therefore, a non-horizontal end-to-end position seems to be an inherent feature of the fibre dynamics on a relatively short time scale, taking into account usually a random initial perturbation of the fibre shape. 

To estimate the rotation along $y$-axis, we proceed as follows.
For each time, we evaluate the moment of inertia tensor and its eigenvectors and eigenvalues. 
The eigenvector corresponding to the largest eigenvalue determines a unit vector $\mathbf{n}$. Using spherical coordinates with vertical zenith direction, vector $\mathbf{n}$ can be determined by the polar angle $\theta$ it makes with the $y$-axis and the azimuthal angle $\phi$ that its horizontal projection makes with the $x$-axis. We evaluate $\phi$ for each time instant $t$ and plot $\phi$ %
in Fig. \ref{fig:figure_simulations_auxplot}b as a function of  the dimensionless instantaneous vertical coordinate $y_{CM}/d$ of the fibre centre of mass, for the numerical trials shown in Figs.~\ref{fig:figure_simulations_14} and \ref{fig:figure_simulations_24}, %
with the same meaning of the colours. It is clear that the polar angle $\phi$ changes significantly with time, also in agreement with the experiments. In the simulations, we observe that the changes of $\phi$ are often non-monotonic, what is related to shape deformations.

In the simulations, we observe that very flexible fibres typically tend to deform out of plane. To describe this feature quantitatively, we %
determine how far the shapes are from %
``an average” plane that contains the centre of mass of the fibre and is spanned by %
two eigenvectors that correspond to the smaller eigenvalues.
In this plane the fibre would be positioned if its shape was flat. 
With this goal in mind, we evaluate the following time-dependent non-flatness parameter $\sigma$: %
\begin{equation}
  \sigma = 
  \sqrt{ 
  \frac{ 1 }{ N } \sum_{i=1}^{N}  h_i ^{2} ,
  }
  \label{eq:flatness}
\end{equation}
where $h_i$ is the distance of the center of bead $i$ from the plane and the summation is over all the beads.
Obviously, $\sigma=0$ if the fibre is planar.

In Fig.~\ref{fig:figure_simulations_auxplot}c, we plot %
$\sigma/d$ 
versus vertical position of the fibre centre-of-mass, $y_{CM}/d$
for the numerical trials shown in Figs.~\ref{fig:figure_simulations_14} and \ref{fig:figure_simulations_24}, %
with the same meaning of the colours. All the fibres, even if initially flat, become non-planar as they settle down, and their non-flatness is pronounced.

\section{Conclusions}
In this paper, we investigated the dynamics of very flexible filaments sedimenting in a viscous fluid at the Reynolds number much smaller than unity. 
In the experimental study, we used ball-chains and a silicon oil. We also performed numerical simulation of the evolution of very elastic fibres with the elasto-gravitational number $B > 4000$, using the precise {\sc Hydromultipole} numerical codes, based on the multipole expansion of the Stokes equations. We have demonstrated that the dynamics of ball-chains and very elastic filaments are similar.

We observed that the dynamics of shorter and longer ball-chains are different. In the early stage of the evolution, shorter ball-chains form shapes that could be approximately classified as vertically oriented planar U-shapes while longer ball-chains form shapes close to vertically oriented planar W-shapes. These findings   agree with numerical studies of sedimenting elastic filaments carried out earlier \cite{lagomarsino2005hydrodynamic,saggiorato2015conformations,koshakji2023robust}, and also in this paper. %
To the best of our knowledge, W-shapes of very flexible filaments have not been observed in experiments so far.

However, we observed that vertically oriented planar U- and W-shapes do not seem to be stable, 
unlike 
more stiff elastic filaments \cite{manghi2006,schlagberger2005orientation,lagomarsino2005hydrodynamic,saggiorato2015conformations,marchetti2018deformation}.
In the experiments, we found that shorter ball-chains typically rotate; moreover, their end-to-end vectors are not horizontal and sometimes even increase the inclination angle with time. %
Longer ball-chains relatively quickly deform significantly and move out of a vertical plane. These features are also present in our numerical simulations of very elastic fibres with a large value of $B$, providing the choice of initially out-of-plane configurations with a non-horizontal end-to-end vector. 

The experimental observations are limited to relatively short times, owing to the height of the container. Therefore, the numerical simulations presented here were also performed for a comparable range of times and vertical distances.
The numerical study of a longtime behaviour of sedimenting very elastic fibres will be the subject of a separate article, with an emphasis on the identification of attractors of the dynamics: periodic motions or stationary configurations. Those results would extend the class of the solutions shown in Ref. \cite{saggiorato2015conformations}.

Finally, in the experiments, we also studied hydrodynamic interactions between two ball-chains, sedimenting one above the other, initially in approximately perpendicular vertical planes. We have shown that, depending on the initial conditions and length, they can approach each other (so-called attraction) or move away from each other (so-called repulsion). In previous studies, the attraction of elastic fibres at symmetric initial configurations one above the other has been found numerically \cite{llopis2007sedimentation,saggiorato2015conformations}, but not the repulsion: our results are new. 

In the literature, two very elastic fibres with large values of $B$ have been observed numerically to always separate from each other if initially  straight and in the same horizontal  plane: collinear \cite{llopis2007sedimentation} or in a symmetric configuration \cite{bukowicki2019sedimenting}. For moderate values of $B$ and the symmetric initial configurations of two elastic fibres, an attracting  stationary relative configuration has been found, with attraction at larger distances and repulsion at smaller ones \cite{bukowicki2019sedimenting,bukowickiJFM}.
The examples discussed above illustrate the complexity of the sedimentation of multiple flexible objects. This work is just a step towards understanding its basic features.

\section*{Conflicts of interest}
There are no conflicts to declare.

\appendix

\section*{Appendices}

\section{Details of the numerical simulations} %
\label{aA}

\subsection{Straight fibres with a very small random perturbation and their evolution}\label{straight}

It is known from the literature \cite{lagomarsino2005hydrodynamic,saggiorato2015conformations} that it takes a lot of time to observe an out-of-plane instability of a very elastic, almost straight fibre, and in this Appendix, we provide some estimations of how fast it grows.

Assume that a straight elastic fibre at the elastic equilibrium is oriented horizontally or inclined at a certain angle $\gamma$ with respect to the horizontal plane, and perturb randomly  positions of all the beads %
with a maximum amplitude %
$0.0001 d$. Examples of the fibre dynamics for such initial configurations are described below.

Let us start with a short fibre with $N=14$ beads. 
In the case of  $almost$ straight initial configurations, when the fibre is oriented horizontally, the fibre stays \textit{almost} in a vertical plane and \textit{almost} symmetric with respect to reflections in the perpendicular vertical plane %
for a relatively long time. 
The initial straight shape is turned into a U-shape relatively quickly but then the fibre keeps such a shape for a long time. %
Only after that, the fibre stops sedimenting vertically only and starts moving also sideways.
For example, in the case of a relatively flexible fibre with the elasto-gravitational number $B=8,000$, vertical sedimentation is observed until about $230 \tau_{b}$ with the fibre travelling to the depth of about $290d$. 
Then symmetry of the U-shape remains to be present but the fibre starts drifting sideways similar to the intermediate mode reported in Ref.~~\cite{saggiorato2015conformations}. 
The results of simulations for such %
initial conditions cannot qualitatively describe experimental results presented in Fig.~\ref{fig:figure2}. 
The initial inclination of the \textit{almost} straight fibre does not lead to a rotation (a feature seen in the experimental trial (b) in Fig.~\ref{fig:figure2}) as the left-right symmetry of the U-shape is quickly restored. %
A non-horizontal
orientation of the end-to-end vector seen in trial (c) in Fig.~\ref{fig:figure2}, does not occur in the simulations under these initial configurations either.

A similar situation is observed for a longer fibre with $N=24$ beads in an $almost$ straight initial configuration. 
For example, for %
$B=8,000$, the initial straight and horizontal shape is turned into a W-shape %
already at $18 \tau_{b}$ at depth of $20d$. 
It then retains an almost symmetric W-shape until $190 \tau_{b}$ at $215d$ depth. 
Then it turns into a hook-shape preserved until the depth of $522d$ at $380 \tau_{b}$.
Nevertheless, the fibre is still oriented 'almost' vertically all this time: the polar angle $\theta$ for unit vector $\mathbf{n}$ starts to differ from $90^{\circ}$ at $\approx 200 \tau_{b}$ but decreases only to $89.7^{\circ}$ at $380 \tau_{b}$. 
This is similar to the initial stage of the sedimentation process presented in Ref.~~\cite{saggiorato2015conformations} in case of larger values of $B$. %
Similar to the case of a short fibre, the results of simulations for almost straight configurations of longer fibres %
cannot qualitatively describe features seen in the experimental results presented in Fig.~\ref{fig:figure3}. In brief, the evolution of tiny perturbations is too slow. We expect that in the experiment, perturbations are not tiny and not as random as, e.g., in the Brownian motion.

Therefore, in this work, we used different initial configurations, with a small (but not tiny) perturbation from a vertical plane. Details will be given in the next section. %

\subsection{C-shaped and propeller-shaped initial configurations}\label{Cpropeller}

Ultimately, we performed simulations for two families of initial configurations of fibres, described below.

  I. The fibre has a planar {\bf C-shape} constructed in three steps:
  
  (1) Shape of a circular arc, i.e. $C$-planar shape, is created using the following equations,
  \begin{equation}
  \begin{cases}
    x_i = r \cdot \sin \left( \left( i - \frac{N+1}{2} \right) \alpha \right) \\
    y_i = 0 \\
    z_i = r \cdot \cos \left( \left( i - \frac{N+1}{2} \right) \alpha \right) - r,
  \end{cases}
  \label{eq:c-shape}
\end{equation}
\\
where $\alpha$ is the curvature angle, $r = l_0 / \left( 2 \cdot \sin \left( \alpha/2 \right) \right) $ is the radius of the circular arc, and $\bm{r}_i=(x_i,y_i,z_i)$.

(2) The whole fibre is then inclined out of the horizontal plane by a certain tilt angle $\gamma$, 
 via the rotation matrix 
\begin{equation}
    \mathcal{R} =
    \begin{bmatrix}
        \cos \gamma & -\sin \gamma & 0 \\
        \sin \gamma &  \cos \gamma & 0 \\
                  0 &            0 & 1
    \end{bmatrix}.
    \label{eq:rot-matrix}
\end{equation}

(3) Positions of all the beads are randomly disturbed with a maximum amplitude of 0.0001$d$.

 II. The fibre has a {\bf propeller shape} constructed in five steps: 
 
 (1) $S$-planar shape is created using Eq.~\ref{eq:s-shape}; 
 \begin{equation}
  \begin{cases}
    x_i = r \cdot \sin \left( \left( i - \frac{2J+1}{2} \right) \alpha \right) \\
    y_i = 0 \\
    z_i = \pm \left( r \cdot \cos \left( \left( i - \frac{2J+1}{2} \right) \alpha \right) - r \right) .
  \end{cases}
  \label{eq:s-shape}
\end{equation}
Here, 
sign ``minus” is taken for the bead number $i$ with $i \le J$ and sign ``plus” is taken for other beads.

 (2) A specific bead $J$ is chosen that, together with the neighbouring bead $(J+1)$ in case of an even $N$, will initially serve as a tip of the propeller.
 
 (3) One arm of the fibre formed by the beads with $i \le J$ is then inclined at a certain angle $\gamma$.
 
 (4) Another arm of the fibre formed by the beads with $i \ge J+1$ is %
  inclined at the angle $-\gamma$.
  
  (5) Positions of beads are randomly disturbed with a maximum amplitude of 0.0001$d$.

As the initial conditions in the simulations shown in Fig.~\ref{fig:figure_simulations_14}, we took shapes with the following values of the parameters: 
\begin{itemize}
    \item C-shape with $\alpha=1^{\circ}$ and $\gamma=1^{\circ}$, %
\item symmetric propeller shape with 
    equal arms ($J\!=\!N/2$),  $\alpha=1^{\circ}$ and $\gamma=5.65^{\circ}$. %
\end{itemize}

As the initial conditions in the simulations shown in Fig.~\ref{fig:figure_simulations_24}, we took shapes with the following values of the parameters: 
\begin{itemize}
    \item C-shape with $\alpha=1^{\circ}$ and $\gamma=1^{\circ}$, %
    \item asymmetric propeller shape with unequal arms ($J\!=\!N/2-1$), $\alpha=1^{\circ}$ and $\gamma=5.65^{\circ}$. %
\end{itemize}

In Figs.~\ref{fig:figure_simulations_14}a-b and Fig.~\ref{fig:figure_simulations_24}a-b, the first rows of the snapshots show the projections of the initial shapes defined above.

\subsection{Evolution of the maximum  bending angle of an elastic fibre} %

In the experiments, the ball-chain local bending angles $\beta_i$ typically did not exceed $33^{\circ}-40^{\circ}$, being smaller than the maximum bending angle out of the fluid (around $55^{\circ}$). In the numerical simulations of flexible fibres, the maximum bending angles were typically of a comparable magnitude as in the experiments. 

In the simulations of a short fibre with $N=14$ beads, for the blue case, the bending angle starts with $1^{\circ}$ at $y_{CM}=0d$, then the maximum bending angle of $57^{\circ}$ is achieved at depths of $y_{CM} \approx 60d$, 
and further down at $y_{CM}=300d$ it decreases to $33^{\circ}$. 
For the red case, the bending angle starts with $6^{\circ}$ at $y_{CM}=0d$, then the maximum bending angle of $44^{\circ}$ is achieved at depths of $y_{CM} \approx 65d$, 
and further down at $y_{CM}=300d$ it decreases to $39^{\circ}$.

In the simulations of a long fibre with $N=24$ beads, for the black case, the bending angle starts with $1^{\circ}$ at $y_{CM}=0d$, then the maximum bending angle of $39^{\circ}$ is achieved at depths of $y_{CM} \approx 100d$, 
and further down at $y_{CM}=300d$ it decreases to $26^{\circ}$.
For the green case, the bending angle starts with $6^{\circ}$ at $y_{CM}=0d$, then the maximum bending angle of $35^{\circ}$ is achieved at depths of $y_{CM} \approx 65d$, 
and further down at $y_{CM}=300d$ it decreases to $29^{\circ}$.

\section{Experiment: description of the videos}

A list of the experimental videos that are available in Electronic supplementary information (ESI)\dag{}. The names of video files refer to the appropriate figure numbers.

\textbf{Video 4a:}
Settling of a single 12-bead ball-chain under gravity in a silicon oil recorded from two cameras with perpendicular lines of sight. Snapshots from this trial are shown in 
Fig.~\ref{fig:figure2}(a). %
The duration of the experiment in real-time is 132s.

\textbf{Video 4b:}
Settling of a single 12-bead ball-chain under gravity in a silicon  oil recorded from two cameras with perpendicular lines of sight. Snapshots from this trial are shown in 
Fig.~\ref{fig:figure2}(b). The ball-chain rotates. 
The duration of the experiment in real-time is 128s.

\textbf{Video 4c:}
Settling of a single 12-bead ball-chain under gravity in a silicon  oil recorded from two cameras with perpendicular lines of sight. Snapshots from this trial are shown in 
Fig.~\ref{fig:figure2}(c).
Rotation and non-horizontal orientation of the end-to-end vector are visible. %
The duration of the experiment in real-time is 136s.

\textbf{Video 5a:}
Settling of a single 20-bead ball-chain under gravity in a silicon  oil recorded from two cameras with perpendicular lines of sight. Snapshots from this trial are shown in 
Fig.~\ref{fig:figure3}(a). Formation of a  W-shape in the early stage of the evolution is visible.
The duration of the experiment in real-time is 110s.

\textbf{Video 5b:}
Settling of a single 20-bead ball-chain under gravity in a silicon  oil recorded from two cameras with perpendicular lines of sight.  Snapshots from this trial are shown in 
Fig.~\ref{fig:figure3}(b).
Formation of a wide, irregular U-shape in the early stage of the evolution is visible.
The duration of the experiment in real-time is 110s.

\textbf{Video 5c:}
Settling of a single 20-bead ball-chain under gravity in a silicon oil recorded from two cameras with perpendicular lines of sight.  Snapshots from this trial are shown in 
Fig.~\ref{fig:figure3}(c). A hook shape is formed very early. 
The duration of the experiment in real-time is 110s.

\textbf{Video 6a:}
Settling of a pair of 12-bead ball-chains under gravity in a silicon oil recorded from two cameras with perpendicular lines of sight.  The ball-chains were placed initially one above the other in perpendicular orientations. Snapshots from this trial are shown in 
Fig.~\ref{fig:figure4}(a). The ball-chains approach each other. 
The duration of the experiment in real-time is 104s.

\textbf{Video 6c:}
Settling of a pair of 12-bead ball-chains under gravity in a silicon oil recorded from two cameras with perpendicular lines of sight. The ball-chains were placed initially one above the other in perpendicular orientations. Snapshots from this trial are shown in 
Fig.~\ref{fig:figure4}(c). The ball-chains move away from each other.
The duration of the experiment in real-time is 108s.

\textbf{Video 7a:}
Settling of a pair of 20-bead ball-chains under gravity in a silicon oil recorded from two cameras with perpendicular lines of sight.  The ball-chains were placed initially one above the other in perpendicular orientations. Snapshots from this trial are shown in 
Fig.~\ref{fig:figure5}(a). The ball-chains approach each other. 
The duration of the experiment in real-time is 108s.

\textbf{Video 7c:}
Settling of a pair of 20-bead ball-chains under gravity in a silicon oil recorded from two cameras with perpendicular lines of sight.  The ball-chains were placed initially one above the other in perpendicular orientations. Snapshots from this trial are shown in 
Fig.~\ref{fig:figure5}(c). The ball-chains move away from each other. 
The duration of the experiment in real-time is 140s.

\section*{Acknowledgements}
This work was supported in part by the National Science Centre under
grant UMO-2018/31/B/ST8/03640. 

\bibliography{references.bib} %

\bibliographystyle{IEEEtran} %

\end{document}